\documentclass[%
 aip,
 amsmath,amssymb,
 reprint,%
]{revtex4-1}

\usepackage{graphicx}
\usepackage{dcolumn}
\usepackage{bm}

\usepackage[utf8]{inputenc}
\usepackage[T1]{fontenc}
\usepackage{mathptmx}
\usepackage{etoolbox}
\usepackage{color}
\usepackage[percent]{overpic}
\definecolor{green2}{RGB}{0,115,4}
\usepackage{chemformula}
\usepackage[version=4]{mhchem}
\usepackage{xr}
\usepackage{amsthm}
\usepackage{hyperref}
\usepackage[english]{babel}

\newtheorem{thm}{Theorem}[section]

\newtheorem{prop}[thm]{Proposition}

\numberwithin{equation}{section}

\DeclareMathOperator{\tr}{tr}

\makeatletter
\def\@email#1#2{%
 \endgroup
 \patchcmd{\titleblock@produce}
  {\frontmatter@RRAPformat}
  {\frontmatter@RRAPformat{\produce@RRAP{*#1\href{mailto:#2}{#2}}}\frontmatter@RRAPformat}
  {}{}
}%
\makeatother
\begin{document}

\title[]{Thermodynamically consistent thixotropic viscoelastic modeling of human blood from a conformational free energy}
\author{Thomas Martinod}
\affiliation{{\'A}rea de Ciencias Fundamentales, 
Escuela de Ciencias Aplicadas e Ingenier{\'\i}a, Universidad EAFIT, Medell{\'\i}n 050022, Colombia}
\author{Andr\'{e}s C\'{o}rdoba}
\email{tmartinods@eafit.edu.co, acordobau@eafit.edu.co}
\affiliation{{\'A}rea de Ciencias Fundamentales, 
Escuela de Ciencias Aplicadas e Ingenier{\'\i}a, Universidad EAFIT, Medell{\'\i}n 050022, Colombia}
\author{Jay D. Schieber}
\affiliation{Pritzker School
of Molecular Engineering, University of Chicago, 
Chicago, Illinois 60637, United States}

\date{\today}

\begin{abstract}

Structural thixotropic viscoelastic models describe the rheology of human blood accurately, but their structural kinetics are postulated, so their compatibility with the second law of thermodynamics is not guaranteed. We present ROUGE (ROUleaux model Guided by free Energy), a tensorial model of blood that describes the deformation of individual red blood cells and of rouleaux with two conformation tensors, and the degree of aggregation with a scalar variable. A single postulated Helmholtz free energy determines the stress, the relaxation of all three variables, and the entropy generation, so the model satisfies the second law in any flow. To build it, we extend a simplified formulation of nonequilibrium thermodynamics, which accounts for conformational energy and entropy, to microstructures described by several structural variables of different kinds. The coupling between aggregation and rouleaux deformation is motivated by the Langmuir adsorption of fibrinogen and is restricted so that the quiescent, fully aggregated state is the stable minimum of the free energy. Fits to published data for step-ups and step-downs in shear rate show that the fibrinogen binding sites remain far from saturation, so that the free energy has no adjustable parameters. They also show that rouleaux breakdown requires no empirical shear-activation term: the structure is destroyed in proportion to the work done by the flow on the aggregates, and a single structural time produces both breakdown within seconds and an algebraic recovery over tens of seconds. The resulting model has five well-determined parameters, and its fitted solvent viscosity coincides with that of plasma.

\par\medskip
\noindent\textit{The following article has been submitted to the Journal of Rheology.}
\end{abstract}

\maketitle

\section{Introduction}\label{sec:intro}

Human blood is a concentrated suspension of cells, mostly red blood cells, in plasma \cite{beris_recent_2021}. At rest and at low shear rates, red blood cells aggregate into stacks known as rouleaux. Aggregation is mediated chiefly by fibrinogen and is reversible: flow breaks the rouleaux apart, and they form again when the flow slows down \cite{baskurt_red_2011}. The rouleaux raise the viscosity at low shear rates, give blood a small apparent yield stress, and add a viscoelastic response that dominates below about $10\,\mathrm{s}^{-1}$. At higher shear rates the individual cells deform and align with the flow, which adds a faster viscoelastic mode and produces marked shear thinning \cite{horner_investigation_2018,horner_measurements_2019,armstrong_methodology_2020}. Blood is therefore a thixo-elasto-visco-plastic (TEVP) material whose stress depends on its deformation history. Describing that history accurately is essential both for simulating flows in the circulation and for using rheology as a diagnostic tool, since aggregation and cell deformability are altered in many pathologies \cite{beris_recent_2021,baskurt_red_2011}.

Generalized Newtonian models, such as the Casson model, describe the steady shear viscosity of blood well \cite{apostolidis_modeling_2014}, but they carry no memory of the flow history. Structural thixotropic models add a scalar structure parameter $\lambda\in[0,1]$, equal to one for fully aggregated blood, whose evolution balances breakdown against aggregation \cite{mewis_thixotropy_2009}. This approach was applied to blood by Apostolidis \textit{et al.}~\cite{apostolidis_modeling_2015}, later enriched with viscoelastic contributions from the cells and the rouleaux \cite{horner_investigation_2018,horner_measurements_2019} and with shear-induced aggregation \cite{armstrong_methodology_2020}, and other TEVP frameworks have also been adapted to blood \cite{saramito_new_2009,wei_multimode_2018,varchanis_modeling_2019,giannokostas_advanced_2020}. The most recent member of this family, the tensorial enhanced thixotropic viscoelastic (t-ETV) model of Armstrong \textit{et al.}~\cite{armstrong_tensorial_2022}, writes the stress as the sum of a White--Metzner--Cross contribution from the cells, a viscoelastic rouleaux contribution with modulus $G_R\lambda^m$, and an elastic contribution associated with the yield stress, while $\lambda$ follows an empirical kinetic law with separate time constants for Brownian aggregation, shear-induced breakdown, and shear-induced aggregation. Fitted to steady and step shear data, the t-ETV predicts small, large, and unidirectional large amplitude oscillatory shear with good accuracy \cite{armstrong_tensorial_2022}.

The structural kinetics of these models, however, are postulated, and their compatibility with the second law of thermodynamics is not guaranteed. Armstrong \textit{et al.} showed that the rouleaux stress of the t-ETV corresponds to a Hookean dumbbell with modulus $G_R\lambda^m$, and left a thermodynamically consistent formulation for future work \cite{armstrong_tensorial_2022}. Completing it requires treating $\lambda$ as a state variable on which the free energy depends, so that its evolution also contributes to the entropy production, and an empirically postulated kinetic law does not by itself keep that contribution nonnegative along arbitrary flow histories, for instance after a sudden reversal of the flow. Thermodynamically grounded models of blood have been derived from polymer network theory \cite{owens_new_2006,moyers-gonzalez_non-homogeneous_2008}, from the bracket formalism \cite{tsimouri_constitutive_2018}, and from population balances \cite{jariwala_application_2020}. To our knowledge, however, none of them retains the structure of the t-ETV, with separate microstructural variables for the cells and the rouleaux and a scalar structure parameter that modulates the rouleaux stress, a structure tested against an extensive set of steady, transient, and oscillatory data \cite{horner_measurements_2019,armstrong_tensorial_2022}.

Nonequilibrium thermodynamics provides ways to build admissibility into a model from the start, most rigorously through the GENERIC formalism \cite{grmela_dynamics_1997,ottinger_dynamics_1997,ottinger_beyond_2005} and the generalized bracket formalism of Beris and Edwards \cite{beris_thermodynamics_1994}; within GENERIC, \"Ottinger~\cite{ottinger_modeling_2002} constructed a general class of models with a tensor and a scalar as structural variables. The mathematical machinery of these formalisms is considerable, which partly explains why most structural models are still proposed phenomenologically. Schieber and C\'ordoba~\cite{schieber_nonequilibrium_2021} proposed a simpler route: they write the energy and entropy balances with the tools of classical transport phenomena and read the thermodynamic restrictions off the entropy production. The stress must then contain a term built from the gradient of the conformational free energy, and the relaxation of the microstructure must dissipate free energy. The approach does not test the Jacobi identity of the reversible dynamics, but it is well suited to models whose reversible terms are known to satisfy it, and it is much easier to apply. Its examples, however, use either a single conformation tensor or several tensors of the same kind. Blood requires a richer description: two conformation tensors of different physical origin, a scalar that cannot be transported by an upper-convected derivative, and rouleaux that store energy as well as entropy, since aggregation is driven by attractive interactions between cells mediated by plasma proteins.

In this work we take up the suggestion of Armstrong \textit{et al.} and extend the t-ETV into a thermodynamically consistent model, which we call ROUGE, for ROUleaux model Guided by free Energy. ROUGE keeps the level of description of the t-ETV, a conformation tensor for the cells, a second one for the rouleaux, and the structure parameter, motivated by the two viscoelastic modes and the thixotropy observed in blood \cite{horner_investigation_2018,horner_measurements_2019,armstrong_tensorial_2022}. To our knowledge, it is the first application of the formalism of Ref.~\cite{schieber_nonequilibrium_2021} to a structural thixotropic model. A single postulated Helmholtz free energy determines the stress, the relaxation of every microstructural variable, and the entropy generation, so that the second law is satisfied by construction in any flow. The free energy reproduces the rouleaux stress of the t-ETV, and requiring the quiescent, fully aggregated state to be its stable minimum restricts the function that couples aggregation to rouleaux deformation. For this function we adopt a deliberately simple form motivated by the Langmuir adsorption of fibrinogen. To keep the whole description at the level of conformation tensors, the cells are described by a conformation tensor of Oldroyd-B type, and the elastic yield-stress contribution of the rouleaux is left out.

Fitting ROUGE to published step changes in shear rate \cite{armstrong_feb20_2021} leads to several conclusions about blood. The data select the low-coverage limit of the Langmuir coupling, in which the binding sites remain far from saturation and the free energy has no adjustable parameters. Rouleaux breakdown needs neither a dedicated time constant nor the empirical shear-activation term of earlier models: it follows from the coupling between the structure and the deformation of the rouleaux, at a rate proportional to the work done by the flow on the aggregates. A single structural time produces both the breakdown of the structure within seconds and its recovery, which is algebraic rather than exponential and extends over tens of seconds. The final model has five well-determined parameters, and its solvent viscosity, when left free, coincides with that of plasma.

The paper is organized so that it can be read at two levels. Sections~\ref{sec:thermodynamics}--\ref{sec:langmuir} construct the model: Sec.~\ref{sec:thermodynamics} extends the energy and entropy balances of Ref.~\cite{schieber_nonequilibrium_2021} to two conformation tensors and a scalar structure parameter, Sec.~\ref{sec:model} presents ROUGE and its stress and relaxation functions, Sec.~\ref{sec:admissibility} establishes the conditions on the coupling function, and Sec.~\ref{sec:langmuir} introduces the Langmuir coupling and its low-coverage limit. These sections state the main results and leave the derivations and proofs to the appendices. Section~\ref{sec:methods} describes the data and the fitting procedure, and Sec.~\ref{sec:results} presents and discusses the results. Readers interested mainly in what the model predicts for blood can go directly to these two sections, where the equations of the selected model are collected in Table~\ref{tab:equations}. Section~\ref{sec:conclusions} concludes. Appendix~\ref{app:entropy_balance} derives the energy and entropy balances, Appendix~\ref{app:proofs} contains the proofs of the propositions, Appendix~\ref{app:admissibility} the stability analysis of the rest state, Appendix~\ref{app:steady_shear} the small-deformation analysis of steady shear, and Appendix~\ref{app:numerics} the numerical and statistical details.

\section{Nonequilibrium thermodynamics of fluids with conformational entropy and energy}\label{sec:thermodynamics}

We follow the approach of Schieber and C\'ordoba~\cite{schieber_nonequilibrium_2021}, who derive the restrictions imposed by the first and second laws on a microstructural model directly from the balances of energy and entropy. They obtained these restrictions for a single conformation tensor. Here we extend them to a microstructure described by two conformation tensors, $\boldsymbol{c}^C$ for the individual cells and $\boldsymbol{c}^R$ for the rouleaux, together with a scalar structure parameter $\lambda$. Both tensors are symmetric and positive definite and equal to the identity tensor, $\boldsymbol{\delta}$, at equilibrium. We use the sign convention of Ref.~\cite{schieber_nonequilibrium_2021}, in which the total stress is $\boldsymbol{\pi}=p\boldsymbol{\delta}+\boldsymbol{\tau}$, with $p$ the thermodynamic pressure and $\boldsymbol{\tau}$ the extra stress, $(\nabla\boldsymbol{v})_{ij}=\partial_iv_j$, with $\boldsymbol{v}$ the fluid velocity, and $\dot{\boldsymbol{\gamma}}=\nabla\boldsymbol{v}+(\nabla\boldsymbol{v})^\dagger$.

The microstructure can store energy as well as entropy. The total energy per unit volume is therefore
\begin{equation}
\epsilon=\frac{1}{2}\rho v^2+\rho\hat{U}+\rho\hat{u}_c(\boldsymbol{c}^C,\boldsymbol{c}^R,\lambda),
\label{eq:energy_density}
\end{equation}
where $\hat{U}$ is the internal energy per unit mass of the fast degrees of freedom, which remain in local equilibrium and act as a heat bath with a well-defined temperature $T$, and $\hat{u}_c$ is the energy per unit mass stored in the microstructure. Similarly, the total entropy per unit mass is split as
\begin{equation}
\hat{S}_{\rm tot}=\hat{S}(\hat{U},\rho)+\hat{s}_c(\boldsymbol{c}^C,\boldsymbol{c}^R,\lambda),
\label{eq:entropy_split}
\end{equation}
where $\hat{S}$ is the entropy of the heat bath. Combining the energy balance with the local-equilibrium relation of the heat bath, and introducing the entropy flux $\boldsymbol{J}_s=\boldsymbol{q}/T$, with $\boldsymbol{q}$ the heat flux, and the conformational Helmholtz free energy $\hat{f}_c=\hat{u}_c-T\hat{s}_c$, the microstructure enters the entropy balance only through the derivatives of $\hat{f}_c$ and the rates of change of the structural variables [Appendix~\ref{app:entropy_balance}, Eq.~\eqref{eq:entropy_balance_general}].

To proceed, we need the evolution equations of the microstructural variables. Both conformation tensors are transported by the flow through the upper-convected derivative,
\begin{equation}
\frac{D\boldsymbol{c}^A}{Dt}=(\nabla\boldsymbol{v})^\dagger\cdot\boldsymbol{c}^A+\boldsymbol{c}^A\cdot\nabla\boldsymbol{v}+\dot{\boldsymbol{c}}^A_{\rm relax},\qquad A\in\{C,R\},
\label{eq:evolution_c}
\end{equation}
where $\dot{\boldsymbol{c}}^A_{\rm relax}$ describes the relaxation of each microstructure toward equilibrium. The structure parameter is a scalar and cannot be transported in the same way. We couple it to the flow linearly through the rate-of-strain tensor,
\begin{equation}
\frac{D\lambda}{Dt}=\boldsymbol{g}:\dot{\boldsymbol{\gamma}}+\dot{\lambda}_{\rm relax},
\label{eq:evolution_lambda}
\end{equation}
where $\boldsymbol{g}(\boldsymbol{c}^C,\boldsymbol{c}^R,\lambda)$ is a symmetric coupling tensor and $\dot{\lambda}_{\rm relax}$ the relaxation of $\lambda$. Unlike the upper-convected terms in \eqref{eq:evolution_c}, which are known to satisfy the Jacobi identity required by GENERIC, the reversible term $\boldsymbol{g}:\dot{\boldsymbol{\gamma}}$ is not tested by the present approach. Its compatibility with the more rigorous formalisms must therefore be checked separately; the particular coupling tensor adopted in Sec.~\ref{sec:model} satisfies the restrictions that GENERIC imposes on the coupling between a conformation tensor and a scalar~\cite{ottinger_modeling_2002}.

Substituting \eqref{eq:evolution_c} and \eqref{eq:evolution_lambda} into the entropy balance, and using the symmetry of $\boldsymbol{c}^A$ and of $\partial\hat{f}_c/\partial\boldsymbol{c}^A$, which commute for an isotropic free energy (Appendix~\ref{app:entropy_balance}), we obtain
\begin{equation}
\begin{split}
\rho\frac{D\hat{S}_{\rm tot}}{Dt} & +\nabla\cdot\boldsymbol{J}_s \\
& = T\boldsymbol{J}_s\cdot\nabla\left(\frac{1}{T}\right)
\\
& -\frac{1}{2T}\left[\boldsymbol{\tau}+2\rho\left(\boldsymbol{c}^C\cdot\frac{\partial\hat{f}_c}{\partial\boldsymbol{c}^C}+\boldsymbol{c}^R\cdot\frac{\partial\hat{f}_c}{\partial\boldsymbol{c}^R}+\frac{\partial\hat{f}_c}{\partial\lambda}\boldsymbol{g}\right)\right]:\dot{\boldsymbol{\gamma}}
\\
& -\frac{\rho}{T}\left[\frac{\partial\hat{f}_c}{\partial\boldsymbol{c}^C}:\dot{\boldsymbol{c}}^C_{\rm relax}+\frac{\partial\hat{f}_c}{\partial\boldsymbol{c}^R}:\dot{\boldsymbol{c}}^R_{\rm relax}+\frac{\partial\hat{f}_c}{\partial\lambda}\dot{\lambda}_{\rm relax}\right].
\end{split}
\label{eq:entropy_balance}
\end{equation}
The left-hand side contains the accumulation and the net flux of entropy, so the right-hand side is the entropy production, which the second law requires to be nonnegative. Its three lines involve driving forces of different tensorial character. In an isotropic material, the Curie principle excludes coupling between the first line, driven by a vector, and the other two, but it does not exclude couplings between the tensorial and scalar forces of the second and third lines. To our knowledge such couplings have not been examined for structural models, and we follow the usual practice of neglecting them, so that each line must be nonnegative separately~\cite{schieber_nonequilibrium_2021}.

The first line is nonnegative for the anisotropic Fourier law $\boldsymbol{q}=-\boldsymbol{k}\cdot\nabla T$ with a positive semidefinite conductivity tensor $\boldsymbol{k}$, since it then equals $\nabla T\cdot\boldsymbol{k}\cdot\nabla T/T^2$. The second line changes sign under a sudden reversal of the flow, because $\dot{\boldsymbol{\gamma}}$ reverses while the microstructure, which relaxes on a finite time scale, does not. It can therefore remain nonnegative for arbitrary flow histories only if the bracket reduces to a purely viscous contribution. This is how the degeneracy restriction of GENERIC manifests itself in the present formalism: apart from the viscous dissipation, the second line on the right side of \eqref{eq:entropy_balance} must vanish. This fixes the extra stress,
\begin{equation}
\boldsymbol{\tau}=-\eta\dot{\boldsymbol{\gamma}}-2\rho\,\boldsymbol{c}^C\cdot\frac{\partial\hat{f}_c}{\partial\boldsymbol{c}^C}-2\rho\,\boldsymbol{c}^R\cdot\frac{\partial\hat{f}_c}{\partial\boldsymbol{c}^R}-2\rho\,\frac{\partial\hat{f}_c}{\partial\lambda}\boldsymbol{g},
\label{eq:stress_general}
\end{equation}
for which the second line equals $\eta\,\dot{\boldsymbol{\gamma}}:\dot{\boldsymbol{\gamma}}/2T$ and is nonnegative for $\eta\geq0$. For an isotropic free energy, $\boldsymbol{c}^A$ and $\partial\hat{f}_c/\partial\boldsymbol{c}^A$ commute, so \eqref{eq:stress_general} is symmetric. The first two elastic terms have the Kramers form found by Ref.~\cite{schieber_nonequilibrium_2021} for a single tensor. The last one is new: through the coupling tensor $\boldsymbol{g}$, the structure parameter contributes directly to the stress, even though it is a scalar.

The third line in \eqref{eq:entropy_balance} is the entropy generation due to the relaxation of the microstructure, $\dot{\sigma}=\dot{\sigma}_C+\dot{\sigma}_R+\dot{\sigma}_\lambda$, with the entropy generation densities
\begin{equation}
\begin{split}
    \dot{\sigma}_A & =-\frac{\rho}{T}\frac{\partial\hat{f}_c}{\partial\boldsymbol{c}^A}:\dot{\boldsymbol{c}}^A_{\rm relax}\\
\qquad
\dot{\sigma}_\lambda & = -\frac{\rho}{T}\frac{\partial\hat{f}_c}{\partial\lambda}\dot{\lambda}_{\rm relax}.
\end{split}
\label{eq:entropy_generation}
\end{equation}
The second law requires only $\dot{\sigma}\geq0$, but we impose the stronger condition $\dot{\sigma}_C,\dot{\sigma}_R,\dot{\sigma}_\lambda\geq0$, so that each relaxation mechanism dissipates free energy on its own and no ordering between mechanisms needs to be proved. As anticipated by Ref.~\cite{schieber_nonequilibrium_2021}, this condition is satisfied automatically when each relaxation term is driven by the free-energy gradient. More generally, the relaxation fluxes $(\dot{\boldsymbol{c}}^C_{\rm relax},\dot{\boldsymbol{c}}^R_{\rm relax},\dot{\lambda}_{\rm relax})$ could be related to the forces $(\partial\hat{f}_c/\partial\boldsymbol{c}^C,\partial\hat{f}_c/\partial\boldsymbol{c}^R,\partial\hat{f}_c/\partial\lambda)$ through a positive semidefinite mobility matrix that couples the three mechanisms. Such a generalization might be interesting, but we do not consider it here, and take each relaxation term to depend only on its own force.

\begin{prop}\label{prop:dissipation}
Let the relaxation terms be
\begin{equation}
\dot{\boldsymbol{c}}^A_{\rm relax}=-M_A\,\boldsymbol{c}^A\cdot\frac{\partial\hat{f}_c}{\partial\boldsymbol{c}^A},\qquad
\dot{\lambda}_{\rm relax}=-M_\lambda\frac{\partial\hat{f}_c}{\partial\lambda},
\label{eq:antiparallel}
\end{equation}
with scalar mobilities $M_C,M_R,M_\lambda\geq0$ that may depend on the state variables. Then $\dot{\sigma}_C,\dot{\sigma}_R,\dot{\sigma}_\lambda\geq0$ for any differentiable isotropic $\hat{f}_c$.
\end{prop}

The proof is given in Appendix~\ref{app:proofs}. Under \eqref{eq:antiparallel}, therefore, the second law no longer restricts the free energy, only the signs of the mobilities. This is the property on which the model of Sec.~\ref{sec:model} is built.

Finally, the separation between energy and entropy has a consequence that has no counterpart in purely entropic models. As shown by the balance of the internal energy of the heat bath [Appendix~\ref{app:entropy_balance}, Eq.~\eqref{eq:internal_energy_bath}], the work done against the entropic part of the stress heats the fluid immediately, as in the single-tensor case~\cite{schieber_nonequilibrium_2021}, while the work done against its energetic part is first stored in the microstructure and released to the heat bath only as the microstructure relaxes. For blood, the second mechanism corresponds to the energy stored in rouleaux bonds and in deformed aggregates, which is returned on the time scales of structural relaxation rather than instantaneously.

\section{The ROUGE model}\label{sec:model}

Section~\ref{sec:thermodynamics} fixed the form of the evolution equations and showed that, when every relaxation term is driven by the gradient of the free energy, the second law reduces to the nonnegativity of the mobilities. To complete the model we must specify three objects: the coupling tensor $\boldsymbol{g}$ in \eqref{eq:evolution_lambda}, the conformational free energy $\hat{f}_c$, and the mobilities in \eqref{eq:antiparallel}. These are the constitutive hypotheses of ROUGE. Everything else, namely the stress, the relaxation functions, and the entropy generation, follows from them.

For the coupling between the structure parameter and the flow we take
\begin{equation}
\boldsymbol{g}=-\lambda\left(\boldsymbol{c}^R-\boldsymbol{\delta}\right),
\label{eq:g_rouge}
\end{equation}
the simplest symmetric tensor that vanishes when the rouleaux are undeformed and is proportional to the amount of structure present. In simple shear, $\boldsymbol{g}:\dot{\boldsymbol{\gamma}}=-2\lambda\dot{\gamma}\,c^R_{xy}$, so the flow destroys structure at a rate set by the shear deformation of the rouleaux rather than directly by the shear rate as in the t-ETV.

For the free energy we postulate
\begin{equation}
\hat{f}_c=\frac{G_C}{2\rho}\left[\tr\boldsymbol{c}^C-\log|\boldsymbol{c}^C|\right]+\frac{G_R}{2\rho}\left[\varphi(\zeta)-\frac{\lambda^m}{m}\right],
\qquad
\zeta=\lambda\,e^{D_R},
\label{eq:free_energy}
\end{equation}
where $G_C$ and $G_R$ are the moduli of the cells and of the rouleaux, $m>1$ is the exponent of the t-ETV, and
\begin{equation}
D_R=\tr\boldsymbol{c}^R-\log|\boldsymbol{c}^R|-3
\label{eq:DR}
\end{equation}
measures the deformation of the rouleaux. Writing $D_R$ in terms of the eigenvalues $c_i$ of $\boldsymbol{c}^R$ as $\sum_i(c_i-\log c_i-1)$ shows that $D_R\geq0$, with equality only for $\boldsymbol{c}^R=\boldsymbol{\delta}$. The first term in \eqref{eq:free_energy} is the free energy of a Hookean dumbbell~\cite{schieber_nonequilibrium_2021} and describes the elasticity of the individual cells. The term $-\lambda^m/m$ lowers the free energy as aggregation proceeds and represents the adhesion energy of the rouleaux. The function $\varphi$ couples aggregation to rouleaux deformation through the variable $\zeta$, which combines the amount of structure with its deformation and equals one in the quiescent, fully aggregated state $\lambda=1$, $\boldsymbol{c}^R=\boldsymbol{\delta}$.

In the t-ETV, the exponent $m$ was introduced as a fitting parameter of the rouleaux modulus $G_R\lambda^m$ and fixed at $m=3/2$ for all donors~\cite{armstrong_tensorial_2022,horner_measurements_2019}, and no physical justification for this value was given. A value $m>1$ makes the elasticity of the aggregates grow faster than the degree of aggregation, which is plausible if the rouleaux also connect into larger networks as they form, but we do not attempt to derive it here. In ROUGE, $m$ acquires an additional, thermodynamic role: it is the exponent of the adhesion energy in \eqref{eq:free_energy}, so it shapes the free-energy landscape of the structure and enters the admissibility conditions of Sec.~\ref{sec:admissibility}, which for the couplings of Sec.~\ref{sec:langmuir} restrict it to $1<m<2$. To keep the comparison with the t-ETV direct, we retain $m=3/2$ throughout.

At this stage $\varphi$ is an arbitrary differentiable function; Secs.~\ref{sec:admissibility} and~\ref{sec:langmuir} constrain it and propose specific forms. It is convenient to work with the dimensionless free energy
\begin{equation}
F=\frac{2\rho}{G_R}\hat{f}_c=\beta\left[\tr\boldsymbol{c}^C-\log|\boldsymbol{c}^C|\right]+\varphi(\zeta)-\frac{\lambda^m}{m},
\qquad
\beta=\frac{G_C}{G_R}.
\label{eq:F}
\end{equation}

The particular combination $\zeta=\lambda e^{D_R}$ is chosen so that the stress reproduces the rouleaux stress of the t-ETV, whatever the coupling function. The derivatives of $F$ that enter the stress and the relaxation functions are
\begin{align}
\nonumber
&\boldsymbol{c}^C\cdot\frac{\partial F}{\partial\boldsymbol{c}^C}=\beta\left(\boldsymbol{c}^C-\boldsymbol{\delta}\right),
\qquad
\boldsymbol{c}^R\cdot\frac{\partial F}{\partial\boldsymbol{c}^R}=\varphi'(\zeta)\,\zeta\left(\boldsymbol{c}^R-\boldsymbol{\delta}\right), \qquad \\ 
&\frac{\partial F}{\partial\lambda}=\varphi'(\zeta)\frac{\zeta}{\lambda}-\lambda^{m-1},
\label{eq:F_derivatives}
\end{align}
which follow from $\partial D_R/\partial\boldsymbol{c}^R=\boldsymbol{\delta}-(\boldsymbol{c}^R)^{-1}$.

\begin{prop}\label{prop:stress}
For any differentiable $\varphi$, the stress \eqref{eq:stress_general} with the coupling tensor \eqref{eq:g_rouge} and the free energy \eqref{eq:free_energy} is
\begin{equation}
\boldsymbol{\tau}=-\eta_C\dot{\boldsymbol{\gamma}}-G_C\left(\boldsymbol{c}^C-\boldsymbol{\delta}\right)-G_R\lambda^m\left(\boldsymbol{c}^R-\boldsymbol{\delta}\right),
\label{eq:stress_rouge}
\end{equation}
where $\eta_C$ is the viscosity of the plasma and of the fast degrees of freedom of the cells.
\end{prop}

The proof is given in Appendix~\ref{app:proofs}. The rouleaux stress in \eqref{eq:stress_rouge} is that of the t-ETV~\cite{armstrong_tensorial_2022}. It arises from two separate contributions: the elasticity of the deformed rouleaux and the change of free energy associated with the flow-induced change of structure. Their sum is independent of $\varphi$, which therefore cannot be identified from the stress alone and must be constrained by other means (Sec.~\ref{sec:admissibility}). The cellular stress is of Oldroyd-B type and replaces the White--Metzner--Cross description of the t-ETV, and the elastic yield-stress contribution of the t-ETV is absent.

The mobilities complete the model. We take
\begin{equation}
M_C=\frac{2\rho}{G_C\tau_C},
\qquad
M_R=\frac{2\rho}{G_R\tau_R\,\zeta^2},
\qquad
M_\lambda=\frac{2\rho}{G_R}\,k_\lambda,
\label{eq:mobilities}
\end{equation}
where $\tau_C$ and $\tau_R$ are relaxation times of the cells and of the rouleaux and $k_\lambda\geq0$ is a rate function, specified in Sec.~\ref{sec:langmuir}, that may depend on the state. The factor $\zeta^{-2}$ in $M_R$ deserves comment. With a constant mobility, the relaxation rate of $\boldsymbol{c}^R$ would be $\varphi'(\zeta)\zeta/\tau_R$, which, for the couplings considered below, grows at least linearly with $\zeta$. Since $\zeta$ grows exponentially with $D_R$, strongly deformed rouleaux would then relax unphysically fast, and the equations would become very stiff. The factor $\zeta^{-2}$ removes this growth while keeping $M_R$ nonnegative, as the second law requires.

Substituting \eqref{eq:F_derivatives} and \eqref{eq:mobilities} into \eqref{eq:antiparallel} gives the relaxation functions of ROUGE,
\begin{align}
\nonumber
&\dot{\boldsymbol{c}}^C_{\rm relax}=\frac{1}{\tau_C}\left(\boldsymbol{\delta}-\boldsymbol{c}^C\right),
\qquad
\dot{\boldsymbol{c}}^R_{\rm relax}=\frac{r(\zeta)}{\tau_R}\left(\boldsymbol{\delta}-\boldsymbol{c}^R\right), \\
\qquad
&\dot{\lambda}_{\rm relax}=-k_\lambda\left[\varphi'(\zeta)\frac{\zeta}{\lambda}-\lambda^{m-1}\right],
\label{eq:relaxations_rouge}
\end{align}
with $r(\zeta)=\varphi'(\zeta)/\zeta$. The cells relax at a constant rate, as in an Oldroyd-B fluid. The rouleaux relax at a rate $r(\zeta)/\tau_R$ that depends on both the amount of structure and its deformation and, as shown in Sec.~\ref{sec:admissibility}, equals $1/\tau_R$ at rest. This rate replaces the relaxation time of the t-ETV, which depends on $\lambda$ and on $\dot{\lambda}$, and follows here from the free energy. Combining \eqref{eq:evolution_lambda}, \eqref{eq:g_rouge}, and \eqref{eq:relaxations_rouge}, the structure parameter evolves as
\begin{equation}
\frac{D\lambda}{Dt}=-\lambda\left(\boldsymbol{c}^R-\boldsymbol{\delta}\right):\dot{\boldsymbol{\gamma}}-k_\lambda\left[\varphi'(\zeta)\frac{\zeta}{\lambda}-\lambda^{m-1}\right],
\label{eq:lambda_rouge}
\end{equation}
where the first term describes breakdown by the flow and the second the relaxation of the structure toward the minimum of the free energy.

The entropy generation densities \eqref{eq:entropy_generation} take explicit forms in which every factor is nonnegative [Appendix~\ref{app:proofs}, Eq.~\eqref{eq:entropy_generation_rouge}]. ROUGE is therefore thermodynamically admissible for any coupling function $\varphi$ and any nonnegative $k_\lambda$. Its parameters are $G_C$, $\tau_C$, $\eta_C$, $G_R$, $\tau_R$, and $m$, together with those introduced by $\varphi$ and $k_\lambda$.

\section{Admissibility conditions on the coupling function}\label{sec:admissibility}

By Proposition~\ref{prop:dissipation}, ROUGE satisfies the second law for any coupling function $\varphi$, as long as $k_\lambda\geq0$. The second law alone therefore leaves $\varphi$ free. Further conditions arise, however, from requiring the free energy to describe blood at rest correctly: the quiescent, fully aggregated state $\boldsymbol{c}^C=\boldsymbol{c}^R=\boldsymbol{\delta}$, $\lambda=1$ must be an equilibrium of the free energy, and that equilibrium must be stable. We state these conditions here; their derivation is given in Appendix~\ref{app:admissibility}.

The cellular part of $F$ is the free energy of a Hookean dumbbell and has its unique minimum at $\boldsymbol{c}^C=\boldsymbol{\delta}$, so only the dependence on $\boldsymbol{c}^R$ and $\lambda$ needs to be examined. Since $\boldsymbol{c}^R$ enters $F$ only through $\zeta=\lambda e^{D_R}$, which increases monotonically with $D_R$ at fixed $\lambda$, the undeformed state $D_R=0$ minimizes $F$ at every $\lambda$ provided that
\begin{equation}
\varphi'(\zeta)\geq0\qquad\forall\,\zeta>0.
\tag{C1}
\end{equation}
Condition (C1) also has a direct dynamical meaning: it is equivalent to $r(\zeta)\geq0$, so by \eqref{eq:relaxations_rouge} it guarantees that the rouleaux relax toward their undeformed state rather than away from it. A negative $\varphi'$ would not violate the second law, since the relaxation would still follow the free-energy gradient; it would describe rouleaux that lower their free energy by deforming, which is unphysical.

For the rest state to be an equilibrium, the thermodynamic force on every variable must vanish there. The force on $\boldsymbol{c}^R$ vanishes at $\boldsymbol{c}^R=\boldsymbol{\delta}$ for any $\varphi$, whereas the force on $\lambda$, given by \eqref{eq:F_derivatives}, equals $\varphi'(1)-1$ at rest. Stationarity therefore requires
\begin{equation}
\varphi'(1)=1.
\tag{C2}
\end{equation}
The mobility $k_\lambda$ introduced in Sec.~\ref{sec:langmuir} vanishes at $\lambda=1$, so the rest state is a fixed point of \eqref{eq:lambda_rouge} even without (C2). We nevertheless require it, so that the equilibrium structure is set by the free energy rather than by the choice of mobility. Condition (C2) also fixes the relaxation rate of the rouleaux at rest: $r(1)=\varphi'(1)=1$, so $\boldsymbol{c}^R$ relaxes with the time constant $\tau_R$ near equilibrium, as anticipated in Sec.~\ref{sec:model}.

For the equilibrium to be locally stable, $F$ must increase in every direction away from it. At rest, the Hessian of $F$ is block diagonal, its $\boldsymbol{c}^R$ block is positive definite by (C2), and $\partial^2F/\partial\lambda^2=\varphi''(1)-(m-1)$ (Appendix~\ref{app:admissibility}). Local stability therefore reduces to
\begin{equation}
\varphi''(1)>m-1.
\tag{C3}
\end{equation}

Local stability does not exclude a second minimum of $F$ at some $\lambda<1$, separated from the rest state by a barrier, which would allow blood to remain trapped in a partially aggregated state at rest. To rule it out, we require the force on $\lambda$ to point toward $\lambda=1$ over the whole physical range when the rouleaux are undeformed, that is,
\begin{equation}
\varphi'(\lambda)<\lambda^{m-1}\qquad\forall\,\lambda\in(0,1).
\tag{C4}
\end{equation}
Because both sides of (C4) coincide at $\lambda=1$ by (C2), condition (C4) implies $\varphi''(1)\geq m-1$, so it contains (C3) except in the borderline case of equality. Together, these conditions make the rest state the global minimum of the free energy.

\begin{prop}\label{prop:global_min}
If $\varphi$ satisfies (C1), (C2), and (C4), the rest state $\boldsymbol{c}^C=\boldsymbol{c}^R=\boldsymbol{\delta}$, $\lambda=1$ is the global minimum of $F$ over all positive definite $\boldsymbol{c}^C$ and $\boldsymbol{c}^R$ and all $\lambda\in(0,1]$.
\end{prop}

In summary, the second law is satisfied by ROUGE for any $\varphi$, and the conditions derived here are conditions on the physical content of the free energy. Condition (C1) ensures that rouleaux relax toward their undeformed state, (C2) makes the rest state an equilibrium of the free energy, (C3) makes it locally stable, and (C4) makes it the global minimum. A coupling function is admissible if it satisfies all four. Whether (C4) imposes restrictions beyond (C3) depends on the form of $\varphi$ and must be checked case by case, as we do in the next section.

\section{Langmuir coupling and its low-coverage limit}\label{sec:langmuir}

Sections~\ref{sec:model} and~\ref{sec:admissibility} leave two functions free in ROUGE: the coupling function $\varphi$ and the structural rate function $k_\lambda$. Any $\varphi$ that satisfies (C1)--(C4), combined with any nonnegative $k_\lambda$, gives a model that reproduces the rouleaux stress of the t-ETV, satisfies the second law, and has a stable rest state. ROUGE is therefore a family of models, and many proposals for $\varphi$ and $k_\lambda$ are possible within it. Here we make a deliberately simple choice. Its purpose is to test whether the framework can describe transient blood rheology with an admissible but otherwise generic free energy, not to provide a definitive microscopic description of aggregation. A more specific model would treat $\lambda$ as a statistical-mechanical variable, for instance as the fraction of bonded cell surface, derive the free energy from the counting of the corresponding microstates, and identify $\varphi$ by comparison with \eqref{eq:F}. We leave that program for future work.

The conditions of Sec.~\ref{sec:admissibility} already exclude the simplest candidates. The choice $\varphi\equiv0$ violates (C2), since $\varphi'(1)=0$. The linear and logarithmic choices $\varphi=\zeta$ and $\varphi=\log\zeta$ satisfy (C1) and (C2), but have $\varphi''(1)=0$ and $\varphi''(1)=-1$, respectively, and therefore violate (C3) for any $m>1$. An admissible coupling must have a nondecreasing derivative with enough curvature at rest, $\varphi''(1)>m-1$.

A form with these properties is suggested by the mechanism of aggregation. Rouleaux are held together by plasma macromolecules, chiefly fibrinogen, bound to sites on the cell surface \cite{baskurt_red_2011}, and the canonical description of the fraction $\theta$ of occupied sites is the Langmuir isotherm. Treating $\zeta$ as the effective variable that drives binding, we write $\theta(\zeta)=\zeta/(\zeta_c+\zeta)$, where $\zeta_c>0$ plays the role of a dissociation constant, measured in the same units as $\zeta$ and therefore directly comparable with its rest value $\zeta=1$. We postulate that $\varphi'$ is proportional to the occupied fraction, $\varphi'(\zeta)=A\,\theta(\zeta)$. Condition (C2) fixes $A=\zeta_c+1$, and integrating with $\varphi(0)=0$ gives
\begin{align}
\nonumber
&\varphi'(\zeta)=\frac{(\zeta_c+1)\,\zeta}{\zeta_c+\zeta}, \\
\qquad
&\varphi(\zeta)=(\zeta_c+1)\left[\zeta-\zeta_c\log\left(1+\frac{\zeta}{\zeta_c}\right)\right].
\label{eq:langmuir_phi}
\end{align}
Condition (C1) holds for every $\zeta_c>0$, and condition (C3) can be satisfied only for $m<2$, in which case it requires
\begin{equation}
\zeta_c>\frac{m-1}{2-m}.
\label{eq:langmuir_condition}
\end{equation}
For the value $m=3/2$ of the t-ETV, this gives $\zeta_c>1$. Condition (C4) turns out to impose nothing further (Appendix~\ref{app:admissibility}): for the Langmuir coupling, local and global stability of the rest state coincide.

With \eqref{eq:langmuir_phi}, the relaxation rate of the rouleaux in \eqref{eq:relaxations_rouge} becomes
\begin{equation}
r(\zeta)=\frac{\zeta_c+1}{\zeta_c+\zeta},
\label{eq:langmuir_r}
\end{equation}
which equals one at rest, is bounded, and decreases with $\zeta$: rouleaux that are more aggregated or more deformed relax more slowly. The thermodynamic force on the structure parameter becomes
\begin{equation}
\frac{\partial F}{\partial\lambda}=\lambda^{m-1}\left[\frac{(\zeta_c+1)\,\lambda^{2-m}e^{2D_R}}{\zeta_c+\lambda e^{D_R}}-1\right].
\label{eq:langmuir_dFdl}
\end{equation}

For the rate function we take
\begin{equation}
k_\lambda=\frac{1}{\tau_\lambda}\,\lambda^{1-m}(1-\lambda)\left[1+\left(b\tilde{D}_R\right)^{d/2}\right],
\qquad
\tilde{D}_R=r(\zeta)^2D_R,
\label{eq:k_lambda}
\end{equation}
with $\tau_\lambda>0$ the time scale of structural relaxation and $b,d\geq0$. The factor $(1-\lambda)$ makes $k_\lambda$ a degenerate mobility that vanishes at the upper bound of the structure parameter, as in phase-field models of bounded order parameters~\cite{cahn_cahnhilliard_1996,elliott_cahnhilliard_1996}, and plays the role of the build-up factor $(1-\lambda)$ of structural kinetic models of thixotropy~\cite{mewis_thixotropy_2009}. It makes the fully aggregated state a fixed point of the relaxation and, together with the factor $\lambda$ in the breakdown term $\boldsymbol{g}:\dot{\boldsymbol{\gamma}}$, keeps $\lambda$ within $[0,1]$. The factor $\lambda^{1-m}$ cancels the factor $\lambda^{m-1}$ in \eqref{eq:langmuir_dFdl}, so that the relaxation remains finite as $\lambda\to0$. The term in the square bracket is an optional shear-activation factor, designed to reproduce the shear-induced aggregation term $|\tau_{\rm aggr}\dot{\gamma}|^d$ of the t-ETV. In that model the activation depends on the shear rate. A mobility, however, should depend only on the state of the material, so we replace the shear rate by an invariant of the rouleaux deformation, rescaled by $r(\zeta)^2$ to compensate for the state-dependent relaxation rate of $\boldsymbol{c}^R$. In steady simple shear at small Weissenberg number $\mathrm{Wi}=\tau_R\dot{\gamma}$, $\tilde{D}_R=\mathrm{Wi}^2+O(\mathrm{Wi}^4)$ for any $\lambda$ and $\zeta_c$ (Appendix~\ref{app:steady_shear}), so the activation factor reduces to the t-ETV term with
\begin{equation}
\tau_{\rm aggr}=\sqrt{b}\,\tau_R.
\label{eq:tau_aggr}
\end{equation}
With \eqref{eq:langmuir_dFdl} and \eqref{eq:k_lambda}, the relaxation of the structure parameter takes the closed form
\begin{equation}
\dot{\lambda}_{\rm relax}=\frac{1-\lambda}{\tau_\lambda}\left[1+\left(b\tilde{D}_R\right)^{d/2}\right]\left[1-\frac{(\zeta_c+1)\,\lambda^{2-m}e^{2D_R}}{\zeta_c+\lambda e^{D_R}}\right].
\label{eq:langmuir_relax}
\end{equation}

The Langmuir family contains a particularly simple member. In the low-coverage limit $\zeta_c\to\infty$, the binding sites are far from saturation, the isotherm reduces to its linear Henry regime, and \eqref{eq:langmuir_phi} tends to $\varphi(\zeta)=\zeta^2/2$, which is finite thanks to the normalization imposed by (C2). The free energy then becomes
\begin{equation}
F=\beta\left[\tr\boldsymbol{c}^C-\log|\boldsymbol{c}^C|\right]+\frac{\lambda^2e^{2D_R}}{2}-\frac{\lambda^m}{m},
\label{eq:henry_F}
\end{equation}
with no adjustable parameters beyond $\beta$ and $m$. It satisfies (C1)--(C4) for every $1<m<2$, since $\varphi'(1)=\varphi''(1)=1$ and $\lambda<\lambda^{m-1}$ on $(0,1)$. In this limit $r\equiv1$, so the rouleaux relax at the constant rate $1/\tau_R$. This is the base relaxation rate of the t-ETV, now derived from the free energy rather than postulated, and without the additional term that the t-ETV introduces during structure buildup~\cite{armstrong_tensorial_2022}. Without activation, the structural relaxation reduces to
\begin{equation}
\dot{\lambda}_{\rm relax}=\frac{1-\lambda}{\tau_\lambda}\left[1-\lambda^{2-m}e^{2D_R}\right].
\label{eq:henry_relax}
\end{equation}

These choices define a hierarchy of nested models, which we compare against data in Sec.~\ref{sec:results}. The most general uses the Langmuir coupling with activation and adds $\zeta_c$, $\tau_\lambda$, $b$, and $d$ to the parameters of Sec.~\ref{sec:model}. Setting $b=0$ removes the activation and leaves $\zeta_c$ and $\tau_\lambda$. Taking in addition $\zeta_c\to\infty$ gives the low-coverage limit, in which the only structural parameter is $\tau_\lambda$. In all cases the stress is the rouleaux stress of the t-ETV, and the entropy generation is nonnegative by construction.

\section{Data and numerical methods}\label{sec:methods}

\subsection{Experimental data}

We use the shear rheology data of human blood published by Armstrong \textit{et al.}~\cite{armstrong_tensorial_2022,armstrong_feb20_2021}, corresponding to their donor 3, with a hematocrit of $42.2\%$. No new measurements were performed for this work. The data were collected with a strain-controlled ARES-G2 rheometer (TA Instruments) in a double-wall Couette geometry at $37\,^\circ\mathrm{C}$, following the handling and measurement protocol of Horner \textit{et al.}~\cite{horner_investigation_2018,horner_measurements_2019}, with a preshear of $300\,\mathrm{s^{-1}}$ for $30\,\mathrm{s}$ before each test to erase the deformation history of the sample~\cite{armstrong_tensorial_2022}. The dataset contains steady-shear, step shear-rate, and small, large, and unidirectional large amplitude oscillatory shear experiments. We fit ROUGE only to the six step experiments: three step-ups from $0.1\,\mathrm{s^{-1}}$ to $\dot\gamma_0=5$, $10$, and $20\,\mathrm{s^{-1}}$, and the three corresponding step-downs from $\dot\gamma_0$ back to $0.1\,\mathrm{s^{-1}}$. Each record lasts about $63\,\mathrm{s}$ and is sampled at logarithmically spaced times, for a total of $N=390$ stress measurements. Unlike the fits of the t-ETV~\cite{armstrong_tensorial_2022}, which combine steady and transient data, ours use transient data only.

\subsection{Simple shear flow and simulated protocol}

In simple shear, $\boldsymbol{v}=(\dot\gamma(t)\,y,0,0)$, and the model reduces to a system of nine ordinary differential equations for the in-plane and $zz$ components of the two conformation tensors and for the structure parameter, given in Appendix~\ref{app:numerics}. With the sign convention of Sec.~\ref{sec:thermodynamics}, the shear stress measured by the rheometer is $-\tau_{xy}$, which by Eq.~\eqref{eq:stress_rouge} is
\begin{equation}
-\tau_{xy}=\eta_C\dot\gamma+G_C\,c^C_{xy}+G_R\lambda^m c^R_{xy}.
\label{eq:shear_stress}
\end{equation}
This is the quantity compared with the data and shown in the figures.

\begin{figure}
\centering
\includegraphics[width=\columnwidth]{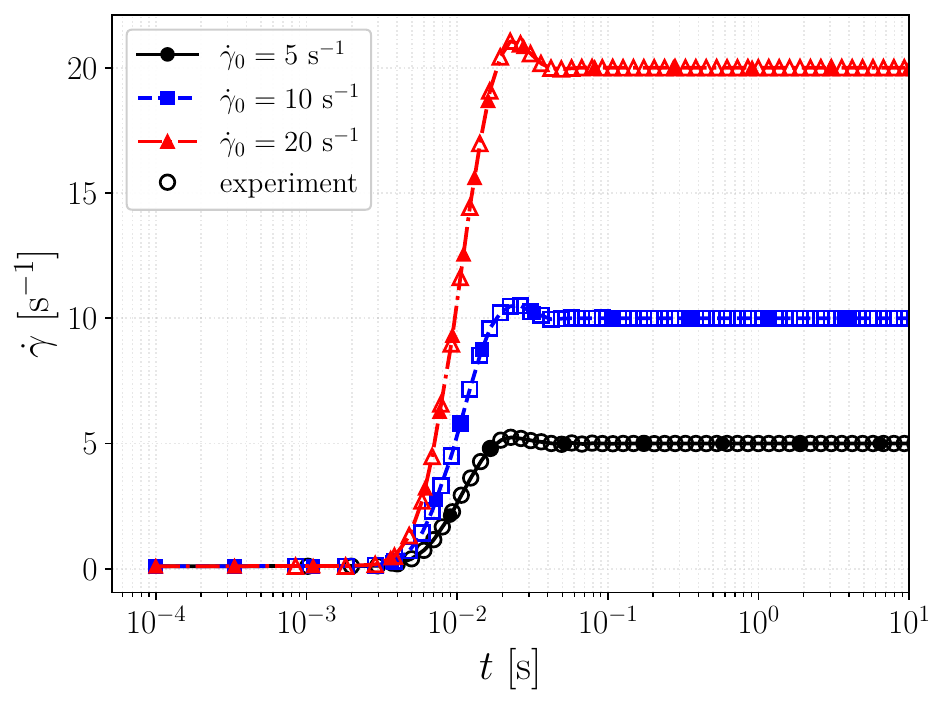}\\
\includegraphics[width=\columnwidth]{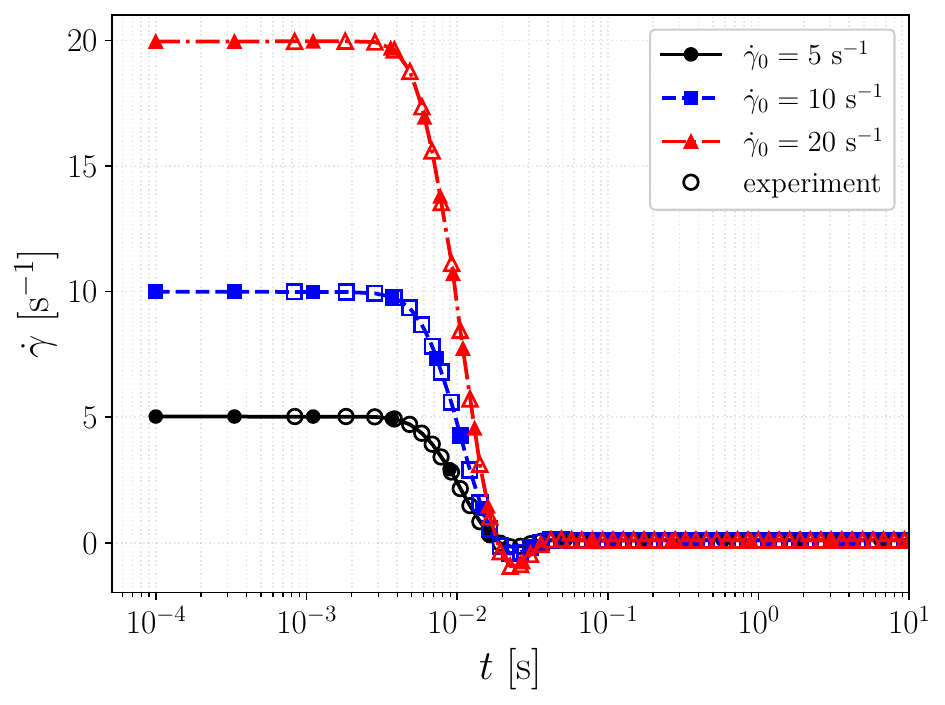}
\caption{Measured shear-rate histories used as input to the simulations: (a, top) step-ups from $0.1\,\mathrm{s^{-1}}$ to $\dot\gamma_0=5$, $10$, and $20\,\mathrm{s^{-1}}$; (b, bottom) step-downs from the same rates to $0.1\,\mathrm{s^{-1}}$. Open symbols, rheometer data~\cite{armstrong_feb20_2021}; lines with filled symbols, linear interpolation imposed in the model. The finite motor response, the overshoot after the step-up, and the brief negative undershoot after the step-down, which reaches about $-1\,\mathrm{s^{-1}}$ near $20\,\mathrm{ms}$ for $\dot\gamma_0=20\,\mathrm{s^{-1}}$, are retained in the simulations.}
\label{fig:rate_histories}
\end{figure}

The model is driven by the measured shear-rate history, interpolated linearly between the recorded times, rather than by an ideal step. Figure~\ref{fig:rate_histories} shows why this matters. The imposed rate takes about $10\,\mathrm{ms}$ to reach its nominal value, overshoots it by a few percent after each step-up, and briefly becomes negative after each step-down. Because the flow-induced breakdown term in Eq.~\eqref{eq:lambda_rouge} changes sign with $\dot\gamma$, these features affect the early stress response and are retained in the simulations.

The initial state of the step-up experiments is not reported, since the duration of the hold at $0.1\,\mathrm{s^{-1}}$ after the preshear is not given in Ref.~\cite{armstrong_tensorial_2022}. We assume that the sample had reached its steady state at $0.1\,\mathrm{s^{-1}}$ before the step. To construct that state, we start from rest, $\boldsymbol{c}^C=\boldsymbol{c}^R=\boldsymbol{\delta}$ and $\lambda=1$, and integrate for $10\,\mathrm{s}$ at $0.1\,\mathrm{s^{-1}}$. At this rate the Weissenberg number $\tau_R\dot\gamma$ is of order $10^{-3}$ or smaller, so the steady state is very close to the rest state: after the hold, $\lambda$ differs from its steady value at $0.1\,\mathrm{s^{-1}}$ by less than $0.01$, which changes the rouleaux stress by less than $1\%$. Each step-down is then integrated from the final state of the corresponding step-up, which has reached steady shear by the end of its record.

The system is stiff, because the cellular and rouleaux relaxation times, of a few milliseconds, are much shorter than the duration of the experiments. We therefore integrate it with an implicit method and an analytic Jacobian; the solver and its settings are described in Appendix~\ref{app:numerics}.

\subsection{Parameter estimation}

The parameters are estimated by minimizing
\begin{align}
\nonumber
&\Phi(\boldsymbol{\theta})=\frac{1}{2}\sum_{c=1}^{6}\frac{1}{N_c}\sum_{i=1}^{N_c}\left[\frac{\tau^{\rm mod}_{xy,c}(t_i;\boldsymbol{\theta})-\tau^{\rm exp}_{xy,c,i}}{s_c}\right]^2, \\ 
\qquad
&s_c=\max_i\left|\tau^{\rm exp}_{xy,c,i}\right|,
\label{eq:objective}
\end{align}
where $c$ runs over the six experiments, $N_c$ is the number of points in the experiment $c$, $\boldsymbol{\theta}$ is the vector of fitting variables, and $\tau^{\rm exp}_{xy,c,i}$ are the measured shear stresses with the sign convention of Eq.~\eqref{eq:stress_rouge}, that is, minus the values reported by the rheometer. The stresses in the step-up experiments are about an order of magnitude larger than in the step-downs. Normalizing each experiment by its maximum stress, as in the fits of the t-ETV~\cite{armstrong_tensorial_2022}, and by its number of points gives every experiment the same weight.

The exponent is fixed at $m=3/2$, as in the t-ETV~\cite{armstrong_tensorial_2022}. The temperature does not affect the stress and enters only the entropy generation, for which we take $T=310\,\mathrm{K}$; the density cancels from all the results. The remaining parameters are fitted in logarithmic form within the bounds of Table~\ref{tab:fit_variables} (Appendix~\ref{app:numerics}). The coupling constant is parametrized so that condition~\eqref{eq:langmuir_condition}, and with it (C1)--(C4), holds at every iterate, and the activation constant is expressed through $\tau_{\rm aggr}$ [Eq.~\eqref{eq:tau_aggr}], so that it can be compared directly with the t-ETV. The low-coverage limit is implemented exactly, with $\varphi(\zeta)=\zeta^2/2$.

\subsection{Identifiability and model selection}

To assess how well the data determine each parameter, we use the covariance matrix of the fitting variables linearized about the optimum [Appendix~\ref{app:numerics}, Eq.~\eqref{eq:covariance}]. Since the fitting variables are logarithms, the square roots of its diagonal entries are relative errors, and its normalized off-diagonal entries are correlation coefficients. The residuals of the fits are largely systematic rather than random, so these quantities indicate which parameters, or combinations of parameters, are identifiable from the data; they should not be read as confidence intervals.

To compare models with different numbers of fitted parameters, we use the Akaike information criterion [AIC, Eq.~\eqref{eq:aic}], which penalizes each additional parameter against the decrease of the cost it produces. The AIC assumes independent errors, which is not strictly the case here, so we use it only to rank nested models fitted to the same data, and lower values indicate the preferred model.

The models compared are the members of the nested hierarchy of Sec.~\ref{sec:langmuir}, and they are fitted in sequence. We first fit the Langmuir coupling with activation, leaving all parameters free. We then compare a finite coupling constant with the low-coverage limit, and the models with and without activation. Guided by the correlations of these fits, we reparametrize the moduli in terms of the viscosities of the cellular and rouleaux viscoelastic modes, $\eta_p^C=G_C\tau_C$ and $\eta_R=G_R\tau_R$. Finally, we fix the solvent viscosity to the viscosity of plasma at $37\,^\circ\mathrm{C}$, $\eta_C=1.2\,\mathrm{mPa\,s}$, within its normal range of $1.10$--$1.30\,\mathrm{mPa\,s}$~\cite{kesmarky_plasma_2008}, and refit. Each fit starts from the optimum of the preceding one.

\section{Results and discussion}\label{sec:results}

\subsection{Unconstrained fit and identifiability}

We first fit the most general member of the hierarchy of Sec.~\ref{sec:langmuir}, the Langmuir coupling with activation, leaving free the eight variables $G_C$, $\eta_C$, $\tau_C$, $G_R$, $\tau_R$, $\tau_\lambda$, $\tau_{\rm aggr}$, and $\zeta_c$, with $d=1/2$ as in the t-ETV~\cite{armstrong_tensorial_2022}. Starting from initial values of the order of those reported for the t-ETV, the cost decreases from $\Phi=4.04\times10^{-2}$ to $1.77\times10^{-2}$. Two variables end at their bounds. The activation time $\tau_{\rm aggr}$ reaches its lower bound, which switches off the activation factor in Eq.~\eqref{eq:k_lambda}, and $\zeta_c$ reaches its upper bound, moving toward the low-coverage limit. The fit therefore points by itself toward the two simplifications examined below.

The correlation matrix of this fit shows strong correlations between pairs of parameters: $G_C$ and $\tau_C$ ($-0.96$), $G_R$ and $\tau_R$ ($-0.91$), $\eta_C$ and $G_R$ ($-0.91$), $\tau_\lambda$ and $\tau_{\rm aggr}$ ($0.93$), and $\tau_\lambda$ and $\zeta_c$ ($0.89$). The first two indicate that the step data determine the products $\eta_p^C=G_C\tau_C$ and $\eta_R=G_R\tau_R$, the viscosities of the cellular and rouleaux viscoelastic modes, much better than the moduli and relaxation times separately. The last two reflect the interplay between the structural time and the two structural mechanisms that the subsequent fits remove. The full correlation matrices of all fits are given in the supplementary material.

\subsection{Model selection}

Table~\ref{tab:fits} summarizes the sequence of fits described in Sec.~\ref{sec:methods}, and leads to three conclusions.

\begin{table}
\caption{Sequence of fits. $k$ is the number of fitted variables, $\Phi$ the cost \eqref{eq:objective}, and $\Delta\mathrm{AIC}$ the difference with respect to the selected model E2. Fit D is the reparametrization of C1 in terms of $\eta_p^C$ and $\eta_R$ and has the same cost.}
\label{tab:fits}
\begin{ruledtabular}
\begin{tabular}{lcccccc}
Fit & $\zeta_c$ & Activation & $\eta_C$ & $k$ & $10^2\,\Phi$ & $\Delta\mathrm{AIC}$\\
\hline
1  & free\footnotemark[1] & free\footnotemark[1] & free & 8 & $1.7703$ & $6.7$\\
B1 & $2$ & free\footnotemark[1] & free & 7 & $1.8343$ & $18.6$\\
B2 & $\to\infty$ & free\footnotemark[1] & free & 7 & $1.7675$ & $4.1$\\
C1 & $2$ & off & free & 6 & $1.8330$ & $16.3$\\
C2 & $2$ & on, $d$ free\footnotemark[1] & free & 8 & $1.8335$ & $20.4$\\
D  & $2$ & off & free & 6 & $1.8330$ & $16.3$\\
E1 & $2$ & off & plasma & 5 & $1.8508$ & $18.1$\\
E2 & $\to\infty$ & off & plasma & 5 & $1.7670$ & $0$\\
\end{tabular}
\end{ruledtabular}
\footnotetext[1]{$\tau_{\rm aggr}$ at its lower bound; in fit 1, $\zeta_c$ also at its upper bound.}
\end{table}

First, the data favor the low-coverage limit of the Langmuir coupling. Fixing $\zeta_c=2$ (B1) instead of letting $\zeta_c\to\infty$ (B2) raises the AIC by $14.5$, and the same comparison with the solvent viscosity fixed (E1 against E2) gives $18.1$. A finite $\zeta_c$ is also compensated by unphysical values of the other parameters: in B1 the solvent viscosity drops to $\eta_C=2.1\times10^{-4}\,\mathrm{Pa\,s}$, below the viscosity of water at $37\,^\circ\mathrm{C}$, while the rouleaux modulus rises and the structural time shortens to $\tau_\lambda\approx0.09\,\mathrm{s}$.

Second, the shear-activation term is unnecessary. The activation time $\tau_{\rm aggr}$ goes to its lower bound in every fit in which it is free. Switching the activation off (C1) leaves the cost unchanged with two fewer parameters, whereas releasing its exponent (C2) leaves $d$ essentially undetermined, with a relative error above $300\%$. The same conclusion holds in the low-coverage limit: E2, without activation, reaches a lower cost than B2, in which $\tau_{\rm aggr}$ again sits at its bound, with two fewer fitted variables. In ROUGE, the breakdown of the rouleaux under flow is therefore carried entirely by the kinematic coupling $\boldsymbol{g}:\dot{\boldsymbol{\gamma}}$, and the structural relaxation reduces to Eq.~\eqref{eq:henry_relax}.

Third, the solvent viscosity coincides with that of plasma. With $\zeta_c=2$, $\eta_C$ is not identifiable, with relative errors between $210\%$ and $375\%$ and a strong correlation with $\tau_R$ ($0.81$ in fit D). In the low-coverage limit with $\eta_C$ free (B2), the fit returns $\eta_C=1.23\times10^{-3}\,\mathrm{Pa\,s}$ without any constraint, within the normal range of plasma viscosity at $37\,^\circ\mathrm{C}$, $1.10$--$1.30\,\mathrm{mPa\,s}$~\cite{kesmarky_plasma_2008}. We therefore fix $\eta_C=1.2\times10^{-3}\,\mathrm{Pa\,s}$ in the last two fits, which leaves five fitted variables. Fit E2, ROUGE in the low-coverage limit without activation, has the lowest cost and the lowest AIC of the whole sequence and is the model analyzed in the rest of this section.

\subsection{The selected model}

Table~\ref{tab:equations} collects the equations of the selected model, both for general flows and in the simple shear form in which it is compared with the data. Together with the parameters of Table~\ref{tab:params}, they are all that is needed to apply the model and to reproduce the curves shown below.

\begin{table*}
\caption{Equations of ROUGE in the low-coverage limit without activation (fit E2). $\overset{\triangledown}{\boldsymbol{c}}=D\boldsymbol{c}/Dt-(\nabla\boldsymbol{v})^\dagger\cdot\boldsymbol{c}-\boldsymbol{c}\cdot\nabla\boldsymbol{v}$ is the upper-convected derivative. Simple shear: $\boldsymbol{v}=(\dot\gamma(t)\,y,0,0)$ starting from rest, for which $c^A_{yy}=c^A_{zz}=1$ at all times, $A\in\{C,R\}$; a dot denotes a time derivative. The measured shear stress is $-\tau_{xy}$.}
\label{tab:equations}
\begin{ruledtabular}
\begin{tabular}{lll}
 & General flow & Simple shear\\
\hline
Stress &
$\boldsymbol{\tau}=-\eta_C\dot{\boldsymbol{\gamma}}-G_C(\boldsymbol{c}^C-\boldsymbol{\delta})-G_R\lambda^m(\boldsymbol{c}^R-\boldsymbol{\delta})$ &
$-\tau_{xy}=\eta_C\dot\gamma+G_C\,c^C_{xy}+G_R\lambda^m c^R_{xy}$\\[4pt]
Cells &
$\overset{\triangledown}{\boldsymbol{c}}{}^C=-(\boldsymbol{c}^C-\boldsymbol{\delta})/\tau_C$ &
$\dot c^C_{xx}=2\dot\gamma\,c^C_{xy}-(c^C_{xx}-1)/\tau_C$,\quad $\dot c^C_{xy}=\dot\gamma-c^C_{xy}/\tau_C$\\[4pt]
Rouleaux &
$\overset{\triangledown}{\boldsymbol{c}}{}^R=-(\boldsymbol{c}^R-\boldsymbol{\delta})/\tau_R$ &
$\dot c^R_{xx}=2\dot\gamma\,c^R_{xy}-(c^R_{xx}-1)/\tau_R$,\quad $\dot c^R_{xy}=\dot\gamma-c^R_{xy}/\tau_R$\\[4pt]
Structure &
$\dfrac{D\lambda}{Dt}=-\lambda(\boldsymbol{c}^R-\boldsymbol{\delta}):\dot{\boldsymbol{\gamma}}+\dfrac{1-\lambda}{\tau_\lambda}\left(1-\lambda^{2-m}e^{2D_R}\right)$ &
$\dot\lambda=-2\lambda\dot\gamma\,c^R_{xy}+\dfrac{1-\lambda}{\tau_\lambda}\left(1-\lambda^{2-m}e^{2D_R}\right)$\\[8pt]
Rouleaux deformation &
$D_R=\tr\boldsymbol{c}^R-\ln\det\boldsymbol{c}^R-3$ &
$D_R=c^R_{xx}-\ln\!\left[c^R_{xx}-(c^R_{xy})^2\right]-1$\\[4pt]
Entropy generation &
$\dot{\sigma}_C=\dfrac{G_C}{2T\tau_C}Q_C$,\quad $\dot{\sigma}_R=\dfrac{G_R}{2T\tau_R}\lambda^2e^{2D_R}Q_R$ &
$Q_A=c^A_{xx}+\dfrac{1+c^A_{xx}}{c^A_{xx}-(c^A_{xy})^2}-3$\\[8pt]
 &
$\dot{\sigma}_\lambda=\dfrac{G_R}{2T\tau_\lambda}\lambda^{m-1}(1-\lambda)\left(\lambda^{2-m}e^{2D_R}-1\right)^2$ & \\[8pt]
 &
$Q_A=\tr\boldsymbol{c}^A+\tr(\boldsymbol{c}^A)^{-1}-6$ & \\
\end{tabular}
\end{ruledtabular}
\end{table*}

Table~\ref{tab:params} lists the parameters of E2. All five fitted variables are well determined, with relative errors below $17\%$, and the largest correlation, between $\eta_p^C$ and $\tau_\lambda$, is $0.79$. The cellular relaxation time, $\tau_C=0.030\,\mathrm{s}$, coincides with the Cross time constant reported for the same donor with the t-ETV, $0.029\,\mathrm{s}$~\cite{armstrong_tensorial_2022}, even though the two models describe the cells differently. The total cellular viscosity of ROUGE, $\eta_C+\eta_p^C=5.3\times10^{-3}\,\mathrm{Pa\,s}$, lies between the zero- and infinite-shear cellular viscosities of the t-ETV, as expected for an Oldroyd-B description that replaces a shear-thinning viscosity by a constant one over the fitted range of shear rates. The parameters of the two models are otherwise not equivalent, since their rouleaux relaxation and structural kinetics have different forms; in particular, $\tau_\lambda$ does not set the recovery time of the structure in ROUGE, as shown below.

\begin{table}
\caption{Parameters of ROUGE in the low-coverage limit without activation (fit E2). Relative errors are indicative (Sec.~\ref{sec:methods}). The t-ETV values for the same donor~\cite{armstrong_tensorial_2022} are given for reference only, since the parameters of the two models are not equivalent.}
\label{tab:params}
\begin{ruledtabular}
\begin{tabular}{lccc}
Parameter & ROUGE & Rel. error & t-ETV\\
\hline
$\eta_p^C=G_C\tau_C$ (Pa\,s) & $4.13\times10^{-3}$ & $5.3\%$ & --\\
$\tau_C$ (s) & $2.97\times10^{-2}$ & $9.9\%$ & $2.9\times10^{-2}$\\
$\eta_R=G_R\tau_R$ (Pa\,s) & $8.05\times10^{-3}$ & $3.5\%$ & $3.2\times10^{-2}$\footnotemark[1]\\
$\tau_R$ (s) & $5.09\times10^{-3}$ & $8.8\%$ & --\\
$\tau_\lambda$ (s) & $0.265$ & $16.4\%$ & $1.79$\\
$G_C$ (Pa)\footnotemark[2] & $0.139$ & -- & $0.833$\\
$G_R$ (Pa)\footnotemark[2] & $1.58$ & -- & $0.212$\\
$\eta_C$ (Pa\,s), fixed & $1.2\times10^{-3}$ & -- & --\\
$m$, fixed & $3/2$ & -- & $3/2$\\
\end{tabular}
\end{ruledtabular}
\footnotetext[1]{Rouleaux structural viscosity $\mu_R$.}
\footnotetext[2]{Derived from the fitted viscosities and relaxation times.}
\end{table}

\begin{figure}
\centering
\includegraphics[width=\columnwidth]{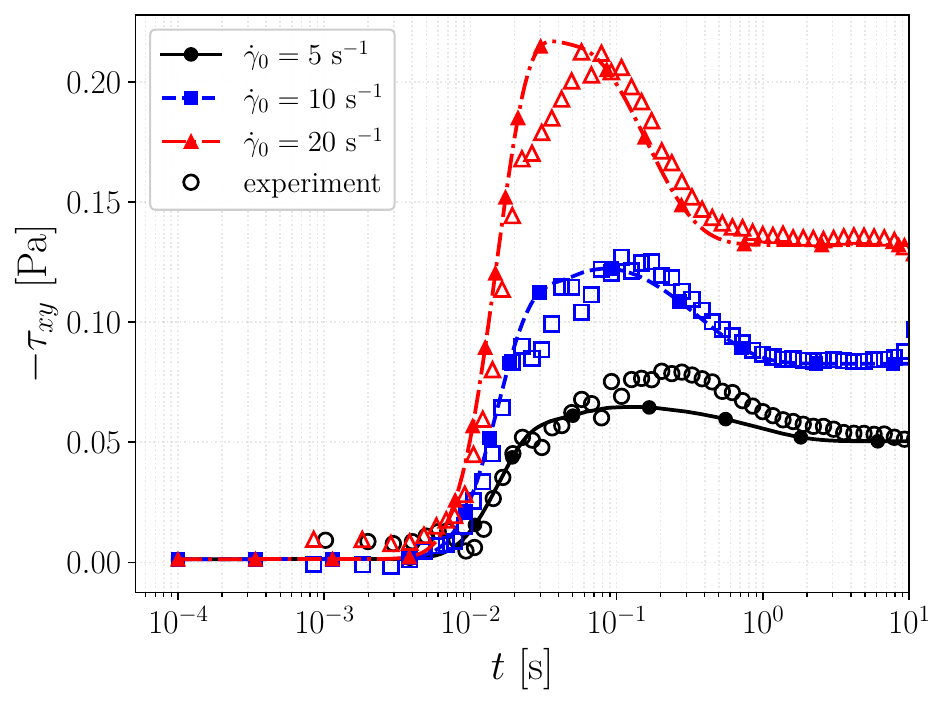}\\
\includegraphics[width=\columnwidth]{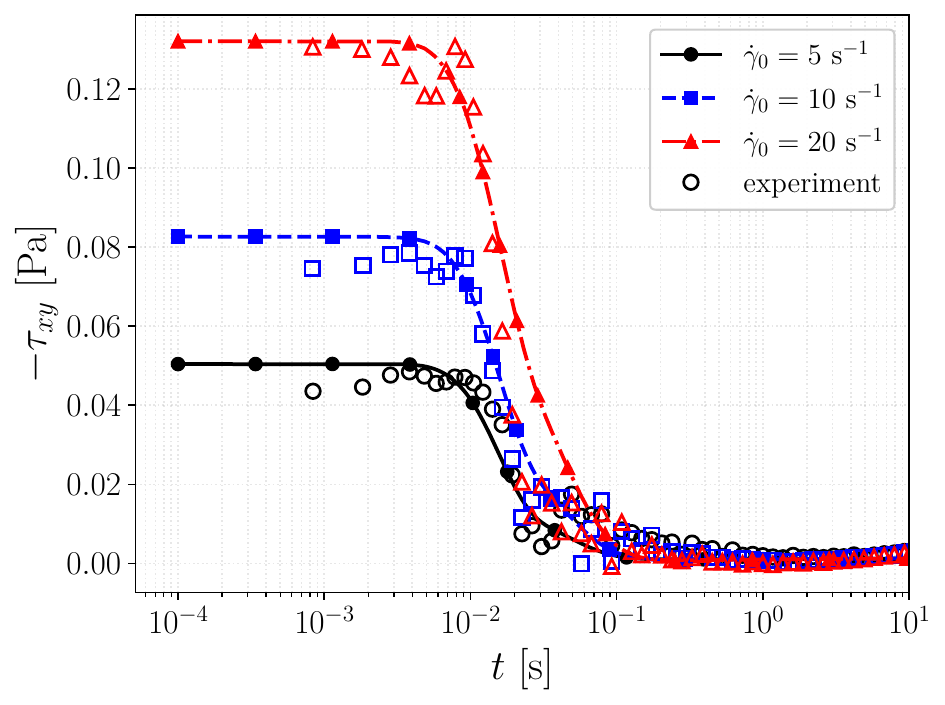}
\caption{Shear stress $-\tau_{xy}$ after (a, top) step-ups from $0.1\,\mathrm{s^{-1}}$ to $\dot\gamma_0=5$, $10$, and $20\,\mathrm{s^{-1}}$ and (b, bottom) step-downs from $\dot\gamma_0$ to $0.1\,\mathrm{s^{-1}}$. Open symbols, data~\cite{armstrong_feb20_2021}; lines with filled symbols, ROUGE in the low-coverage limit with the parameters of Table~\ref{tab:params}.}
\label{fig:stress}
\end{figure}

Figure~\ref{fig:stress} compares the stress predicted by ROUGE with the data. After each step-up [Fig.~\ref{fig:stress}(a)], the stress rises within about $20\,\mathrm{ms}$, following the imposed rate, passes through an overshoot, and decays toward a steady value. The overshoot is the elastic loading of the cells and of the still intact rouleaux, and the subsequent decay reflects the breakdown of the structure. ROUGE reproduces the height and timing of the overshoot at $10$ and $20\,\mathrm{s^{-1}}$, peaking at about $0.22\,\mathrm{Pa}$ near $40\,\mathrm{ms}$ for $20\,\mathrm{s^{-1}}$, but underestimates it at $5\,\mathrm{s^{-1}}$, where the measured peak is about $0.08\,\mathrm{Pa}$ and the predicted one about $0.065\,\mathrm{Pa}$. The predicted steady stresses, $0.050$, $0.083$, and $0.132\,\mathrm{Pa}$, agree with the measured ones a few seconds after the step; at longer times the measured stress keeps decreasing slowly, by about $10\%$ over the following minute, whereas the prediction is already steady. After each step-down [Fig.~\ref{fig:stress}(b)], the stress drops with the shear rate within about $30\,\mathrm{ms}$, closely following the data, and then stays at a few millipascals while the structure re-forms. The per-curve costs (supplementary material) show that the fit is worst at $5\,\mathrm{s^{-1}}$ in both protocols and improves with the shear rate in the step-ups; the step-down from $10\,\mathrm{s^{-1}}$ is the best-described curve. The origin of the remaining discrepancies is discussed in Sec.~\ref{sec:limitations}.

\begin{figure}
\centering
\includegraphics[width=\columnwidth]{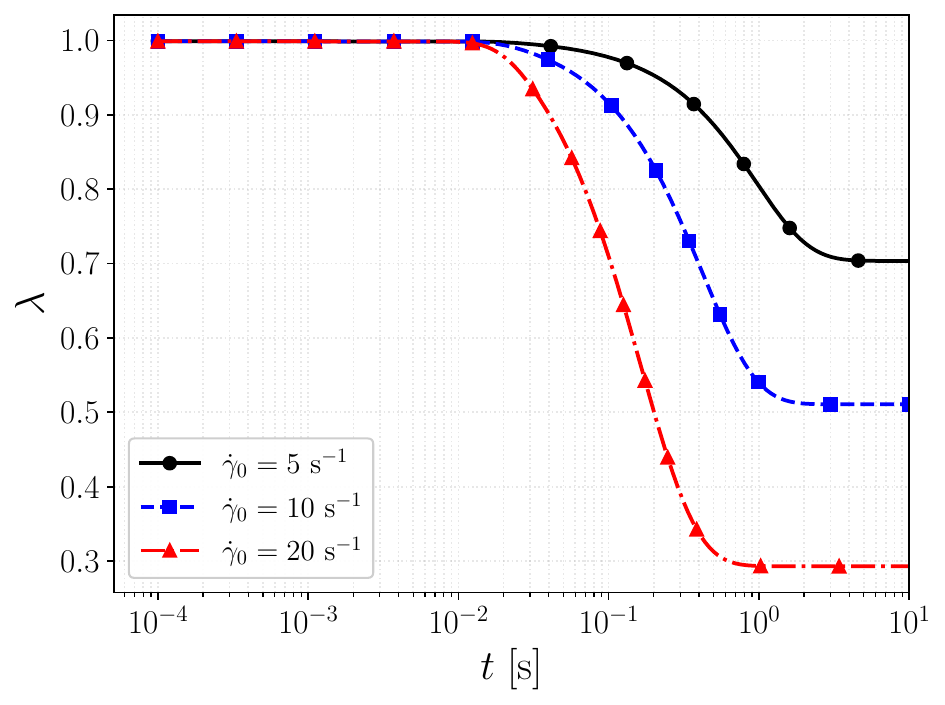}\\
\includegraphics[width=\columnwidth]{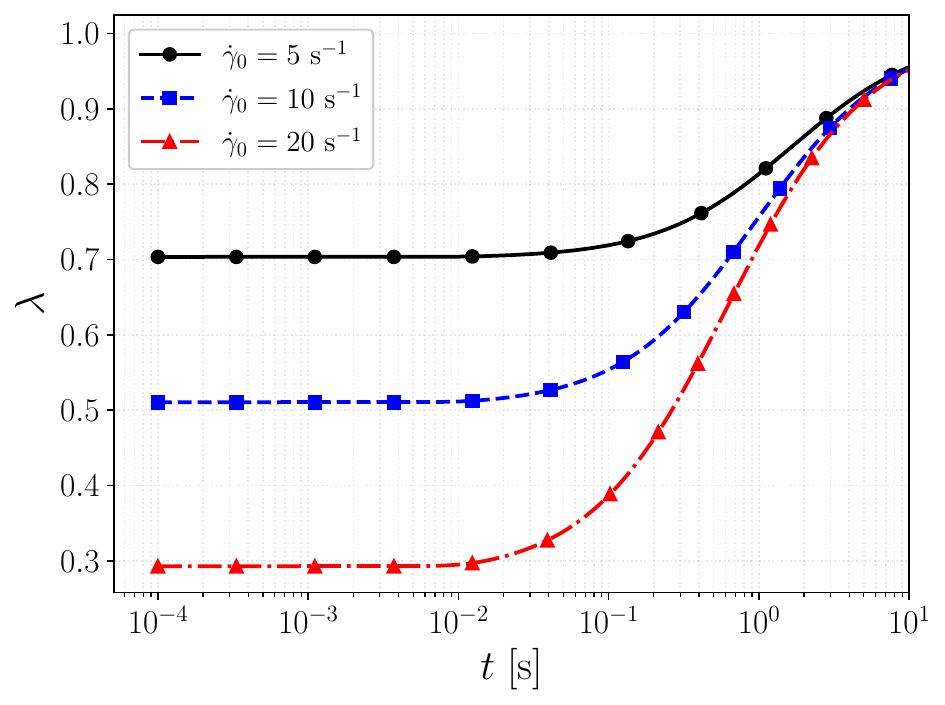}
\caption{Structure parameter during (a, top) the step-ups and (b, bottom) the step-downs of Fig.~\ref{fig:stress}.}
\label{fig:lambda}
\end{figure}

Figure~\ref{fig:lambda} shows the corresponding evolution of the structure parameter. Before the step-up, $\lambda$ is essentially one. After the step, it starts to decrease once the rouleaux have been deformed, at about $15$, $25$, and $55\,\mathrm{ms}$ for $20$, $10$, and $5\,\mathrm{s^{-1}}$, and reaches steady values of $0.293$, $0.511$, and $0.704$ within about $1$ to $3\,\mathrm{s}$. The breakdown has no dedicated time constant. It is driven by the term $-2\lambda\dot\gamma c^R_{xy}$, which builds up as the rouleaux deform, so its onset is delayed with respect to the step, and it is faster at higher rates because, in steady shear at small Weissenberg number, $c^R_{xy}\approx\tau_R\dot\gamma$ and the breakdown rate scales as $\dot\gamma^2$. Comparing Figs.~\ref{fig:stress}(a) and~\ref{fig:lambda}(a), the overshoot occurs while $\lambda$ is still close to one, and the decay of the stress after the overshoot coincides with the breakdown of the structure.

After the step-down [Fig.~\ref{fig:lambda}(b)], the structure starts to recover between about $20$ and $70\,\mathrm{ms}$, the three curves merge after a few seconds, and $\lambda$ reaches $0.95$ at $10\,\mathrm{s}$ and $0.99$ at $50\,\mathrm{s}$. This recovery is much slower than $\tau_\lambda=0.26\,\mathrm{s}$ would suggest, and its form follows from the degenerate mobility. At $0.1\,\mathrm{s^{-1}}$ the rouleaux deformation is negligible, $D_R\approx0$, and near $\lambda=1$ Eq.~\eqref{eq:henry_relax} reduces to
\begin{equation}
\frac{d\lambda}{dt}\approx\frac{2-m}{\tau_\lambda}\,(1-\lambda)^2
\quad\Longrightarrow\quad
1-\lambda\approx\frac{\tau_\lambda}{(2-m)\,t},
\label{eq:algebraic_recovery}
\end{equation}
so the structure approaches full aggregation algebraically rather than exponentially. For $m=3/2$ and the fitted $\tau_\lambda$, Eq.~\eqref{eq:algebraic_recovery} gives $\lambda\approx0.95$ at $10\,\mathrm{s}$ and $0.99$ at $50\,\mathrm{s}$, as in Fig.~\ref{fig:lambda}(b). A single structural time therefore produces both the breakdown within seconds after a step-up and a re-aggregation that extends over tens of seconds after a step-down. The selection of the low-coverage limit has a simple interpretation: over the whole range of structure and deformation probed by these experiments, the fibrinogen binding sites remain far from saturation, and the coupling between aggregation and rouleaux deformation is described by the linear, Henry regime of the isotherm. In this regime the free energy has no parameters beyond $\beta$ and $m$. Moreover, since $\varphi'(\zeta)=\zeta$, the factor $\zeta^2$ in the thermodynamic force on $\boldsymbol{c}^R$ [Eq.~\eqref{eq:F_derivatives}] cancels the factor $\zeta^{-2}$ in the mobility $M_R$ [Eq.~\eqref{eq:mobilities}], so that, although $M_R$ depends on the state, the rouleaux relax with the constant rate $1/\tau_R$.

\subsection{Entropy generation}

\begin{figure}
\centering
\includegraphics[width=\columnwidth]{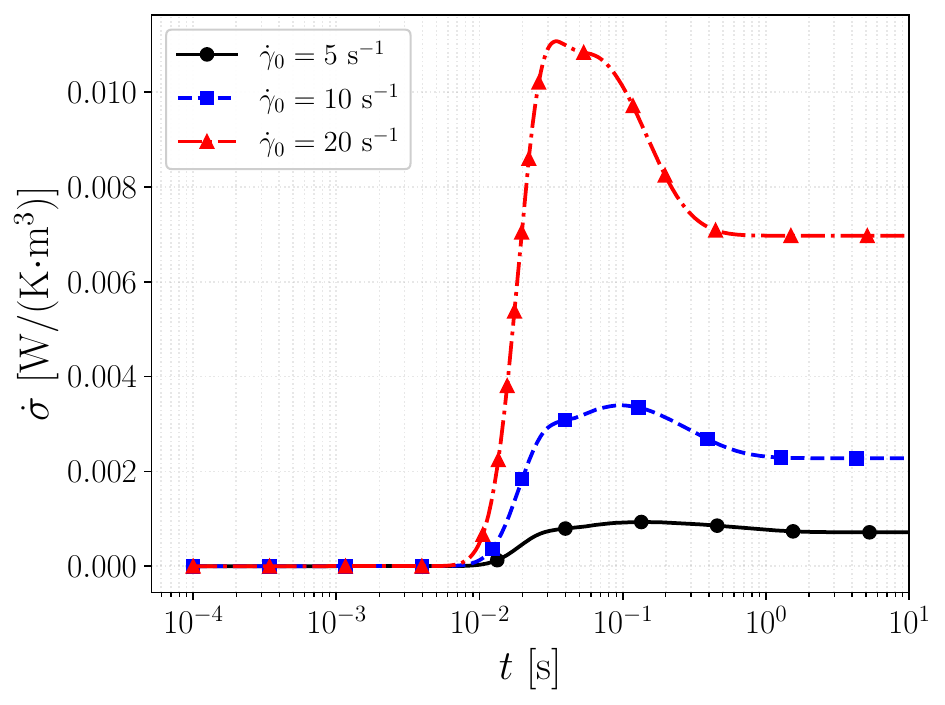}\\
\includegraphics[width=\columnwidth]{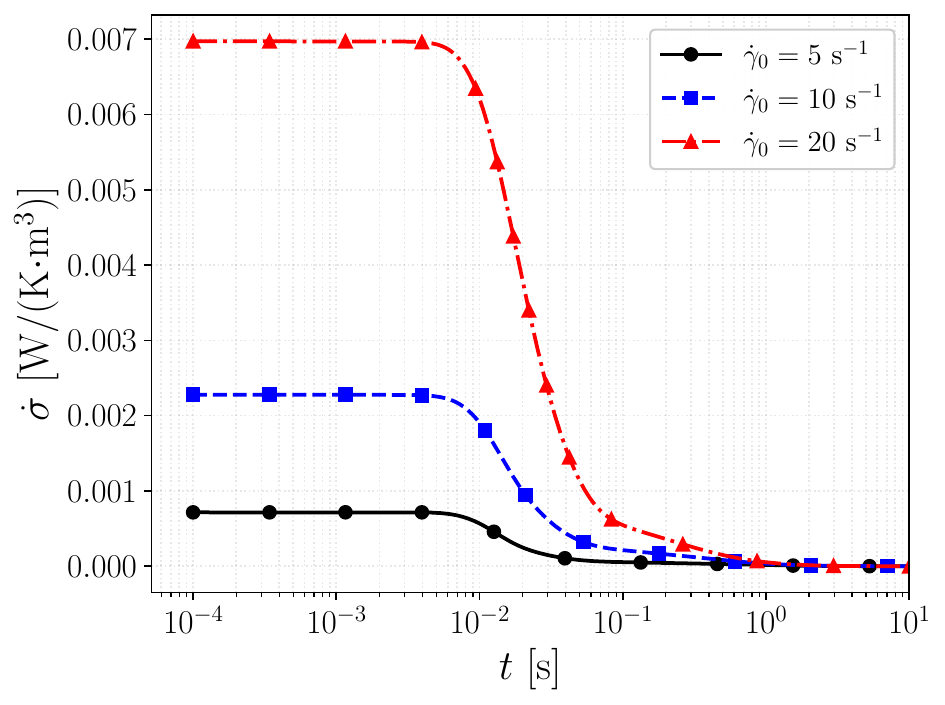}
\caption{Entropy generation by microstructural relaxation, $\dot{\sigma}=\dot{\sigma}_C+\dot{\sigma}_R+\dot{\sigma}_\lambda$ [Eq.~\eqref{eq:entropy_generation_rouge}], during (a, top) the step-ups and (b, bottom) the step-downs of Fig.~\ref{fig:stress}, for $T=310\,\mathrm{K}$.}
\label{fig:entropy}
\end{figure}

Figure~\ref{fig:entropy} shows the entropy generation by microstructural relaxation, $\dot{\sigma}=\dot{\sigma}_C+\dot{\sigma}_R+\dot{\sigma}_\lambda$, computed from Eq.~\eqref{eq:entropy_generation_rouge}. It is nonnegative throughout every protocol, as guaranteed by construction, and the linear scale of the figure makes this directly visible. Before the step-up, at $0.1\,\mathrm{s^{-1}}$, it is negligible on this scale, about $4\times10^{-7}\,\mathrm{W\,K^{-1}\,m^{-3}}$. After the step it rises within about $20\,\mathrm{ms}$ and passes through a maximum of $9.3\times10^{-4}$, $3.4\times10^{-3}$, and $1.1\times10^{-2}\,\mathrm{W\,K^{-1}\,m^{-3}}$ at $5$, $10$, and $20\,\mathrm{s^{-1}}$, between about $35$ and $140\,\mathrm{ms}$ after the step and $30$ to $60\%$ above the steady value. The maximum coincides with the stress overshoot, when the rouleaux are loaded before the structure breaks down. The entropy generation then settles within a few seconds to steady values of $7\times10^{-4}$, $2.3\times10^{-3}$, and $7\times10^{-3}\,\mathrm{W\,K^{-1}\,m^{-3}}$. After the step-down, $\dot{\sigma}$ falls below $10\%$ of its initial value within about $0.1\,\mathrm{s}$ and below $1\%$ within about $2\,\mathrm{s}$, without becoming negative at any time, even while the imposed shear rate briefly changes sign. The viscous dissipation $\eta_C\dot{\boldsymbol{\gamma}}:\dot{\boldsymbol{\gamma}}/2T$ is not included in $\dot{\sigma}$ and is nonnegative independently.

The partition of $\dot{\sigma}$ among the three mechanisms, shown for $10\,\mathrm{s^{-1}}$ in the supplementary material, identifies the origin of each stage. During the step-up, the rouleaux dissipate most in the first tenth of a second, while they are loaded, and $\dot{\sigma}_R$ peaks at about $2.6\times10^{-3}\,\mathrm{W\,K^{-1}\,m^{-3}}$. In steady shear, the cells dominate, with $\dot{\sigma}_C\approx1.3\times10^{-3}$, followed by $\dot{\sigma}_R\approx7\times10^{-4}$ and $\dot{\sigma}_\lambda\approx2.7\times10^{-4}\,\mathrm{W\,K^{-1}\,m^{-3}}$. The structural contribution grows only after about $30\,\mathrm{ms}$, when the breakdown starts, and vanishes momentarily shortly after the step, when the thermodynamic force $\partial F/\partial\lambda$ changes sign. During the step-down, the cellular and rouleaux contributions relax within about $0.2\,\mathrm{s}$, and from then on the dissipation is dominated by $\dot{\sigma}_\lambda$, that is, by the re-aggregation of the structure. This contribution is small, of order $10^{-4}\,\mathrm{W\,K^{-1}\,m^{-3}}$ at $0.1\,\mathrm{s}$ and below $10^{-5}\,\mathrm{W\,K^{-1}\,m^{-3}}$ after a few seconds, so it is barely visible on the scale of Fig.~\ref{fig:entropy}(b), but it persists for tens of seconds and decays with the algebraic recovery~\eqref{eq:algebraic_recovery}.

\begin{figure}
\centering
\includegraphics[width=\columnwidth]{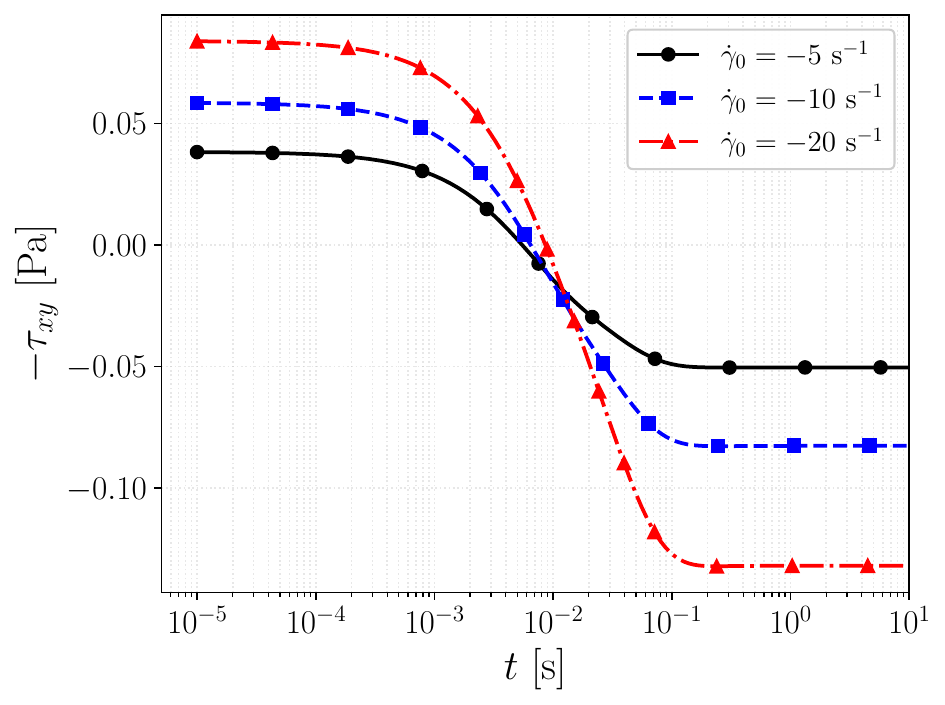}\\
\includegraphics[width=\columnwidth]{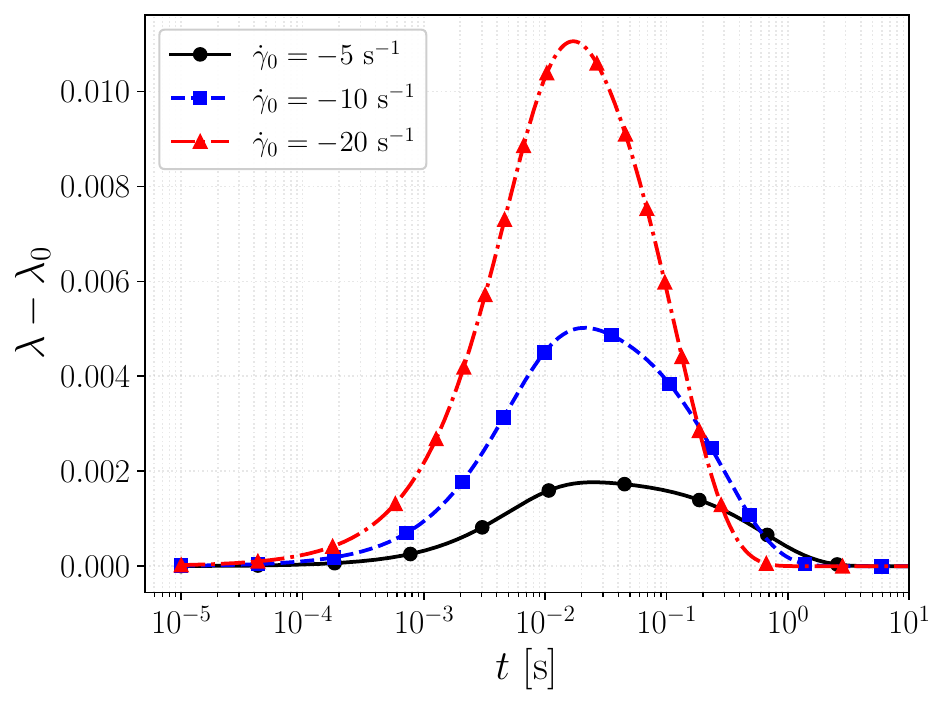}\\
\includegraphics[width=\columnwidth]{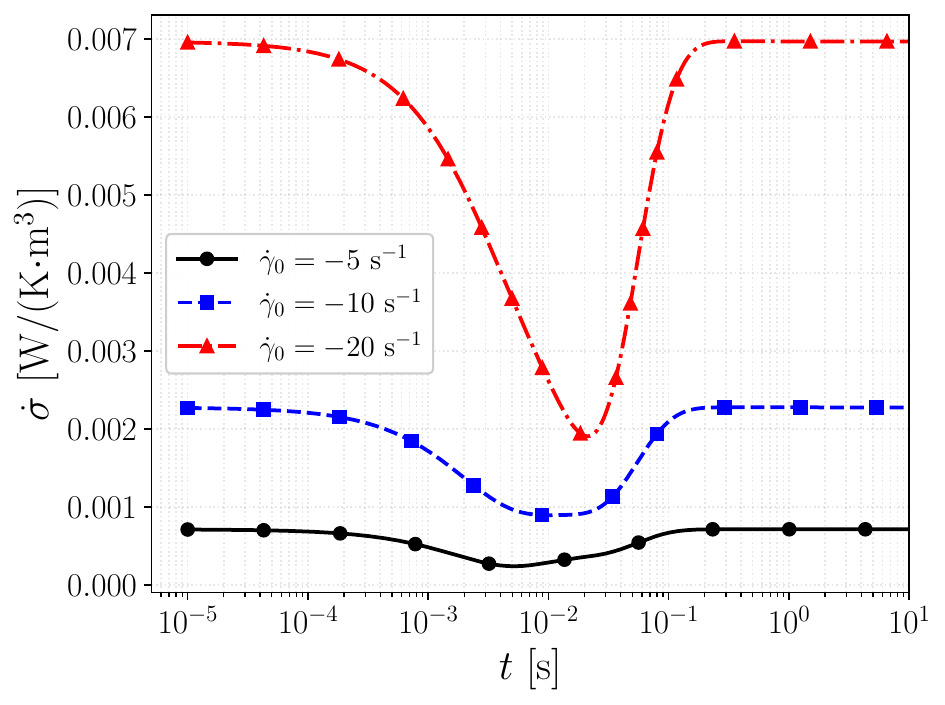}
\caption{Predicted response of ROUGE to a sudden reversal of the flow, from steady shear at $\dot\gamma_0$ to $-\dot\gamma_0$: (a, top) shear stress $-\tau_{xy}$, (b, middle) change of the structure parameter with respect to its steady value $\lambda_0$ before the reversal, and (c, bottom) entropy generation by microstructural relaxation.}
\label{fig:reversal}
\end{figure}

A sudden reversal of the flow is a stringent test of thermodynamic consistency, because the microstructure, oriented by the previous flow, is suddenly sheared in the opposite direction~\cite{schieber_nonequilibrium_2021}. Figure~\ref{fig:reversal} shows the prediction of ROUGE for a reversal from steady shear at $\dot\gamma_0$ to $-\dot\gamma_0$; no data are available for this protocol in the dataset used here. The stress [Fig.~\ref{fig:reversal}(a)] first jumps by $-2\eta_C\dot\gamma_0$, the reversal of the solvent contribution, and then changes sign within about $10\,\mathrm{ms}$ as the cells and the rouleaux are deformed in the opposite direction. It reaches the steady value of opposite sign within about $0.1$ to $0.3\,\mathrm{s}$, without an overshoot, since the structure is already broken down to its steady level. The reversal also reveals a qualitative feature of the kinematic coupling [Fig.~\ref{fig:reversal}(b)]. Immediately after the reversal, $c^R_{xy}$ is still positive while $\dot\gamma$ is negative, so the breakdown term $-2\lambda\dot\gamma c^R_{xy}$ becomes positive and the structure grows transiently. The increase is small, up to about $0.01$ at $20\,\mathrm{s^{-1}}$, peaks a few tens of milliseconds after the reversal, when the rouleaux have reoriented, and disappears within about $1\,\mathrm{s}$. This transient is specific to the kinematic coupling: a kinetic law that depends only on $|\dot\gamma|$, such as that of the t-ETV, leaves $\lambda$ unchanged through the reversal. The entropy generation [Fig.~\ref{fig:reversal}(c)] remains positive throughout. It decreases to about $0.3$ to $0.4$ of its steady value, at a time that increases from about $5\,\mathrm{ms}$ at $5\,\mathrm{s^{-1}}$ to about $20\,\mathrm{ms}$ at $20\,\mathrm{s^{-1}}$, as the microstructure passes through its least deformed configuration, and then returns to the same steady value.

\subsection{Breakdown and recovery compared with the t-ETV}

The ingredients that ROUGE takes over from, adapts, and adds to the t-ETV are summarized in Table~\ref{tab:comparison}. Here we compare the two structural kinetics in light of the results above. In the t-ETV, the structure parameter obeys~\cite{armstrong_tensorial_2022}
\begin{equation}
\frac{D\lambda}{Dt}=\frac{1}{\tau_\lambda}\left[(1-\lambda)-\lambda\,\tau_{\rm break}|\dot\gamma|+(1-\lambda)|\tau_{\rm aggr}\dot\gamma|^d\right],
\label{eq:tetv_kinetics}
\end{equation}
with three time constants for Brownian aggregation, shear-induced breakdown, and shear-induced aggregation. In ROUGE, both the rouleaux relaxation and the structural kinetics follow from the free energy, and the selected model uses a single structural time.

The most significant difference lies in the breakdown mechanism. In the t-ETV, the breakdown rate is proportional to $\lambda|\dot\gamma|$. In ROUGE, it is $-\boldsymbol{g}:\dot{\boldsymbol{\gamma}}=\lambda(\boldsymbol{c}^R-\boldsymbol{\delta}):\dot{\boldsymbol{\gamma}}$, which can be written as
\begin{equation}
-\boldsymbol{g}:\dot{\boldsymbol{\gamma}}=\frac{2\lambda^{1-m}}{G_R}\,\mathcal{W}_R,
\qquad
\mathcal{W}_R=-\frac{1}{2}\boldsymbol{\tau}^R:\dot{\boldsymbol{\gamma}},
\label{eq:breakdown_power}
\end{equation}
where $\boldsymbol{\tau}^R=-G_R\lambda^m(\boldsymbol{c}^R-\boldsymbol{\delta})$ is the rouleaux stress and $\mathcal{W}_R$ the rate of work done by the flow on the rouleaux. Structure is therefore broken in proportion to the elastic work done on the aggregates, without a separate breakdown time. In steady shear at small Weissenberg number, the breakdown rate scales as $2\lambda\tau_R\dot\gamma^2$, quadratically rather than linearly in the shear rate, so the two models predict different dependences of the steady structure on the shear rate at low rates, a difference that steady-shear data could discriminate. The two models also differ in how the structure recovers: in the t-ETV the recovery at low rates is essentially exponential with time $\tau_\lambda$, whereas in ROUGE it is algebraic [Eq.~\eqref{eq:algebraic_recovery}], so the two values of $\tau_\lambda$ in Table~\ref{tab:params} should not be compared as recovery times. Finally, the breakdown term of ROUGE carries the sign of the rouleaux deformation relative to the flow, which produces the transient structure build-up after a reversal shown in Fig.~\ref{fig:reversal}.

\begin{table*}
\caption{Comparison between the t-ETV~\cite{armstrong_tensorial_2022} and ROUGE in the form selected by the fits (E2).}
\label{tab:comparison}
\begin{ruledtabular}
\begin{tabular}{lll}
 & t-ETV & ROUGE\\
\hline
Microstructural variables & Cell stress, rouleaux stress, $\lambda$ & Conformation tensors $\boldsymbol{c}^C$, $\boldsymbol{c}^R$; $\lambda$\\
Cell stress & White--Metzner--Cross (shear thinning) & Oldroyd-B (constant viscosity)\\
Rouleaux stress & $-G_R\lambda^m(\boldsymbol{c}^R-\boldsymbol{\delta})$, postulated & Same, derived from the free energy\\
Elastic yield-stress contribution & Included & Omitted\\
Rouleaux relaxation rate & Depends on $\lambda$ and $\dot\lambda$, postulated & Constant, $1/\tau_R$, derived from the free energy\\
Structural breakdown & $\propto\lambda|\dot\gamma|$, time constant $\tau_{\rm break}$ & $\propto$ work done by the flow on the rouleaux\\
Shear-induced aggregation & $\propto(1-\lambda)|\tau_{\rm aggr}\dot\gamma|^d$ & Not needed\\
Structural time constants & $\tau_\lambda$, $\tau_{\rm break}$, $\tau_{\rm aggr}$ & $\tau_\lambda$\\
Recovery after flow cessation & Approximately exponential & Algebraic, $1-\lambda\propto t^{-1}$\\
Structure after a flow reversal & Unchanged & Small transient build-up\\
Second law & Not guaranteed & Satisfied by construction\\
Data used in the fit & Steady and transient shear & Transient shear (step-ups and step-downs)\\
\end{tabular}
\end{ruledtabular}
\end{table*}

\subsection{Limitations}\label{sec:limitations}

The main discrepancy between ROUGE and the data appears at low shear rates and is associated with the elastic rouleaux stress that ROUGE omits. At $0.1\,\mathrm{s^{-1}}$, with the structure nearly intact, ROUGE predicts a shear stress of about $0.1\,\mathrm{s^{-1}}\times(\eta_C+\eta_p^C+\eta_R)\approx1.3\,\mathrm{mPa}$. The measured stress at this rate, before the step-ups and at the end of the step-downs, is about $4\,\mathrm{mPa}$; after the step-down from $5\,\mathrm{s^{-1}}$, for instance, it rises from about $1.7$ to $4.2\,\mathrm{mPa}$ between $1$ and $63\,\mathrm{s}$. The difference, about $3\,\mathrm{mPa}$, is comparable to the yield stress reported with the t-ETV for the same donor, $\sigma_{y0}=3.6\,\mathrm{mPa}$~\cite{armstrong_tensorial_2022}, which grows with the structure and is carried in that model by the elastic rouleaux stress. Because the data are sampled at logarithmically spaced times, a large fraction of the points of each step-down lies in this low-rate regime, which explains a significant part of their cost. The same missing contribution is likely responsible for the underestimated overshoot at $5\,\mathrm{s^{-1}}$, the lowest rate in the step-ups. Incorporating an elastoviscoplastic rouleaux contribution into the free energy, in a form compatible with Sec.~\ref{sec:thermodynamics}, is the most important extension of the model.

A second discrepancy is the slow decrease of the measured stress over tens of seconds after the step-ups, which ROUGE, whose structure reaches its steady level within a few seconds, does not capture. It may reflect a slower stage of breakdown of the rouleaux network, which would require a second structural variable or a more elaborate rate function $k_\lambda$, or slow processes in the measurement itself; distinguishing between these possibilities requires additional protocols. Both extensions of $k_\lambda$ remain within ROUGE and preserve admissibility as long as $k_\lambda\geq0$.

The cellular contribution is described by an Oldroyd-B model, which captures the fitted range of shear rates but not the shear thinning of individual cells at higher rates; this can be introduced within the framework through a cellular mobility $M_C$ that depends on the deformation of the cells. The exponent $m$ was fixed at the value used in the t-ETV rather than fitted. The model has been fitted to the step experiments of a single donor in homogeneous simple shear, and its predictions for oscillatory and nonhomogeneous flows remain to be tested. Finally, the uncertainties of Table~\ref{tab:params} are indicative, since the residuals of the fits are largely systematic.

\section{Conclusions}\label{sec:conclusions}

We have built ROUGE, a thermodynamically consistent, fully tensorial structural model of human blood, by applying the nonequilibrium thermodynamics of Ref.~\cite{schieber_nonequilibrium_2021} to a microstructure that stores conformational energy as well as entropy and that is described by two conformation tensors, one for the individual cells and one for the rouleaux, together with a scalar structure parameter. The scalar is coupled to the flow through the rate-of-strain tensor, and this coupling contributes directly to the stress. With a single postulated Helmholtz free energy and every relaxation term taken antiparallel to its gradient, the stress, the three relaxation functions, and the entropy generation all follow from the free energy, and the second law reduces to the nonnegativity of the mobilities. The free energy reproduces the rouleaux stress of the t-ETV for any choice of the function that couples aggregation to rouleaux deformation, and requiring the quiescent, fully aggregated state to be its stable global minimum restricts that function through conditions (C1)--(C4). ROUGE is therefore a family of admissible models rather than a single one, in which the coupling function and the structural rate function can be chosen freely within those conditions.

Within this family, we tested a coupling motivated by the Langmuir adsorption of fibrinogen against published step-up and step-down data for human blood. A systematic comparison of nested models showed that the data select the low-coverage limit of the coupling, in which the free energy has no adjustable parameters and the rouleaux relax at the constant rate assumed in the t-ETV, now as a consequence of the free energy. The same comparison showed that the empirical shear-activation term of earlier models is unnecessary: the breakdown of the rouleaux follows from the kinematic coupling alone, at a rate proportional to the work done by the flow on the aggregates. The resulting model has five well-determined parameters, its solvent viscosity coincides with that of plasma when left free, and a degenerate structural mobility with a single time constant reproduces both the breakdown of the structure within seconds and its algebraic recovery over tens of seconds. The entropy generation is nonnegative in every protocol, including a sudden reversal of the flow, for which ROUGE predicts a small transient build-up of structure that kinetic laws based on the magnitude of the shear rate cannot produce.

Several extensions follow directly from this work. The largest discrepancy with the data occurs at low shear rates and points to the elastic rouleaux stress associated with the yield stress, which ROUGE omits; formulating an elastoviscoplastic rouleaux contribution within the same free-energy framework is the next step. The shear thinning of individual cells and a slower stage of structural breakdown can be incorporated through state-dependent mobilities or additional structural variables without affecting admissibility. On the side of the free energy, the Langmuir coupling was chosen as a deliberately simple test; treating $\lambda$ as a statistical-mechanical variable and deriving the free energy from the counting of aggregated microstates would provide a microscopic basis for the coupling function. Finally, on the thermodynamic side, the reversible coupling of the structure parameter to the flow is not tested by the present formalism, but the coupling tensor adopted here satisfies the restrictions that GENERIC imposes on the coupling between a conformation tensor and a scalar~\cite{ottinger_modeling_2002}. The question that remains open is how the energy stored in the rouleaux, released as they re-form, affects the internal energy balance and the heating of blood in nonisothermal flows.

We would like to
dedicate this paper to Arthur B. Metzner. The theoretical 
work presented here is an extension of a model that was 
originally based on his work.
We hope that here we have been able to 
provide an illustration of how 
his work continues to contribute to a deeper understanding 
of the rheology of complex fluids.

\appendix

\section{Energy and entropy balances}\label{app:entropy_balance}

This appendix derives the entropy balance \eqref{eq:entropy_balance} and the balance of the internal energy of the heat bath, following the procedure of Ref.~\cite{schieber_nonequilibrium_2021}.

In the absence of body forces, the total energy balance and the mechanical energy balance read
\begin{align}
\nonumber
&\rho\frac{D}{Dt}\left(\frac{1}{2}v^2+\hat{U}+\hat{u}_c\right)=-\nabla\cdot\boldsymbol{q}-\nabla\cdot(\boldsymbol{\pi}\cdot\boldsymbol{v}), \\
\qquad
&\rho\frac{D}{Dt}\left(\frac{1}{2}v^2\right)=-\boldsymbol{v}\cdot(\nabla\cdot\boldsymbol{\pi}).
\end{align}
Since $\boldsymbol{\pi}$ is symmetric, $\nabla\cdot(\boldsymbol{\pi}\cdot\boldsymbol{v})=\boldsymbol{v}\cdot(\nabla\cdot\boldsymbol{\pi})+\boldsymbol{\pi}:\nabla\boldsymbol{v}$, and subtracting the two balances with $\boldsymbol{\pi}=p\boldsymbol{\delta}+\boldsymbol{\tau}$ gives
\begin{equation}
\rho\frac{D}{Dt}\left(\hat{U}+\hat{u}_c\right)=-\nabla\cdot\boldsymbol{q}-p\,\nabla\cdot\boldsymbol{v}-\boldsymbol{\tau}:\nabla\boldsymbol{v}.
\label{eq:internal_energy_total}
\end{equation}

The fast degrees of freedom are in local equilibrium, so $\hat{S}(\hat{U},\rho)$ obeys the Gibbs relation $T\,d\hat{S}=d\hat{U}+p\,d(1/\rho)$. With the continuity equation, $D\rho/Dt=-\rho\nabla\cdot\boldsymbol{v}$, this becomes
\begin{equation}
\rho T\frac{D\hat{S}}{Dt}=\rho\frac{D\hat{U}}{Dt}+p\,\nabla\cdot\boldsymbol{v}.
\end{equation}
Adding $\rho\,D\hat{s}_c/Dt$ and eliminating $\rho\,D\hat{U}/Dt$ with Eq.~\eqref{eq:internal_energy_total},
\begin{equation}
\rho\frac{D\hat{S}_{\rm tot}}{Dt}=-\frac{1}{T}\nabla\cdot\boldsymbol{q}-\frac{1}{T}\boldsymbol{\tau}:\nabla\boldsymbol{v}-\frac{\rho}{T}\left(\frac{D\hat{u}_c}{Dt}-T\frac{D\hat{s}_c}{Dt}\right).
\end{equation}
Because $\hat{u}_c$ and $\hat{s}_c$ depend only on $x\in\{\boldsymbol{c}^C,\boldsymbol{c}^R,\lambda\}$ and not on $T$, the chain rule gives
\begin{equation}
\frac{D\hat{u}_c}{Dt}-T\frac{D\hat{s}_c}{Dt}=\sum_x\left(\frac{\partial\hat{u}_c}{\partial x}-T\frac{\partial\hat{s}_c}{\partial x}\right)\star\frac{Dx}{Dt}=\sum_x\frac{\partial\hat{f}_c}{\partial x}\star\frac{Dx}{Dt},
\end{equation}
where $\star$ denotes the double contraction for the tensors and the product for $\lambda$, and the derivatives of $\hat{f}_c=\hat{u}_c-T\hat{s}_c$ are taken at fixed $T$. Writing $-\nabla\cdot\boldsymbol{q}/T=-\nabla\cdot\boldsymbol{J}_s+T\boldsymbol{J}_s\cdot\nabla(1/T)$ with $\boldsymbol{J}_s=\boldsymbol{q}/T$, and $\boldsymbol{\tau}:\nabla\boldsymbol{v}=\frac{1}{2}\boldsymbol{\tau}:\dot{\boldsymbol{\gamma}}$, which follows from the symmetry of $\boldsymbol{\tau}$, we obtain
\begin{equation}
\begin{split}
\rho\frac{D\hat{S}_{\rm tot}}{Dt}& + \nabla\cdot\boldsymbol{J}_s\\
& =T\boldsymbol{J}_s\cdot\nabla\left(\frac{1}{T}\right)-\frac{1}{2T}\boldsymbol{\tau}:\dot{\boldsymbol{\gamma}}
\\
& -\frac{\rho}{T}\left[\frac{\partial\hat{f}_c}{\partial\boldsymbol{c}^C}:\frac{D\boldsymbol{c}^C}{Dt}+\frac{\partial\hat{f}_c}{\partial\boldsymbol{c}^R}:\frac{D\boldsymbol{c}^R}{Dt}+\frac{\partial\hat{f}_c}{\partial\lambda}\frac{D\lambda}{Dt}\right].
\end{split}
\label{eq:entropy_balance_general}
\end{equation}

To substitute the evolution equations \eqref{eq:evolution_c} and \eqref{eq:evolution_lambda}, we need the contraction of a symmetric tensor $\boldsymbol{A}$ with the upper-convected terms. Using the symmetry of $\boldsymbol{A}$ and $\boldsymbol{c}^A$ and relabeling indices,
\begin{equation}
\boldsymbol{A}:\left[(\nabla\boldsymbol{v})^\dagger\cdot\boldsymbol{c}^A+\boldsymbol{c}^A\cdot\nabla\boldsymbol{v}\right]=2\left(\boldsymbol{c}^A\cdot\boldsymbol{A}\right):\nabla\boldsymbol{v}.
\end{equation}
For an isotropic free energy, $\boldsymbol{A}=\partial\hat{f}_c/\partial\boldsymbol{c}^A$ is an isotropic function of $\boldsymbol{c}^A$ and commutes with it, so $\boldsymbol{c}^A\cdot\boldsymbol{A}$ is symmetric and $2(\boldsymbol{c}^A\cdot\boldsymbol{A}):\nabla\boldsymbol{v}=(\boldsymbol{c}^A\cdot\boldsymbol{A}):\dot{\boldsymbol{\gamma}}$. This holds for the free energy~\eqref{eq:free_energy}, in which each conformation tensor enters only through its own trace and determinant. Substituting Eqs.~\eqref{eq:evolution_c} and~\eqref{eq:evolution_lambda} into Eq.~\eqref{eq:entropy_balance_general} and collecting the terms proportional to $\dot{\boldsymbol{\gamma}}$ gives Eq.~\eqref{eq:entropy_balance}.

Finally, the balance of $\hat{U}$ follows from Eq.~\eqref{eq:internal_energy_total} written as $\rho\,D\hat{U}/Dt=-\nabla\cdot\boldsymbol{q}-p\nabla\cdot\boldsymbol{v}-\boldsymbol{\tau}:\nabla\boldsymbol{v}-\rho\,D\hat{u}_c/Dt$. Inserting the stress~\eqref{eq:stress_general},
\begin{equation}
-\boldsymbol{\tau}:\nabla\boldsymbol{v}=\frac{\eta}{2}\dot{\boldsymbol{\gamma}}:\dot{\boldsymbol{\gamma}}+\rho\left(\boldsymbol{c}^C\cdot\frac{\partial\hat{f}_c}{\partial\boldsymbol{c}^C}+\boldsymbol{c}^R\cdot\frac{\partial\hat{f}_c}{\partial\boldsymbol{c}^R}+\frac{\partial\hat{f}_c}{\partial\lambda}\boldsymbol{g}\right):\dot{\boldsymbol{\gamma}},
\end{equation}
while the evolution equations give
\begin{align}
\nonumber
\rho\frac{D\hat{u}_c}{Dt}&=\rho\left(\boldsymbol{c}^C\cdot\frac{\partial\hat{u}_c}{\partial\boldsymbol{c}^C}+\boldsymbol{c}^R\cdot\frac{\partial\hat{u}_c}{\partial\boldsymbol{c}^R}+\frac{\partial\hat{u}_c}{\partial\lambda}\boldsymbol{g}\right):\dot{\boldsymbol{\gamma}} \\ 
+&\rho\left(\frac{\partial\hat{u}_c}{\partial\boldsymbol{c}^C}:\dot{\boldsymbol{c}}^C_{\rm relax}+\frac{\partial\hat{u}_c}{\partial\boldsymbol{c}^R}:\dot{\boldsymbol{c}}^R_{\rm relax}+\frac{\partial\hat{u}_c}{\partial\lambda}\dot{\lambda}_{\rm relax}\right).
\end{align}
Subtracting, the flow terms combine through $\partial\hat{f}_c/\partial x-\partial\hat{u}_c/\partial x=-T\,\partial\hat{s}_c/\partial x$, and with Fourier's law $\boldsymbol{q}=-\boldsymbol{k}\cdot\nabla T$ the result is
\begin{equation}
\begin{split}
\rho\frac{D\hat{U}}{Dt}
&=\nabla\cdot(\boldsymbol{k}\cdot\nabla T)-p\,\nabla\cdot\boldsymbol{v}+\frac{\eta}{2}\dot{\boldsymbol{\gamma}}:\dot{\boldsymbol{\gamma}}
\\
& -\rho T\left(\boldsymbol{c}^C\cdot\frac{\partial\hat{s}_c}{\partial\boldsymbol{c}^C}+\boldsymbol{c}^R\cdot\frac{\partial\hat{s}_c}{\partial\boldsymbol{c}^R}+\frac{\partial\hat{s}_c}{\partial\lambda}\boldsymbol{g}\right):\dot{\boldsymbol{\gamma}}
\\
&-\rho\left(\frac{\partial\hat{u}_c}{\partial\boldsymbol{c}^C}:\dot{\boldsymbol{c}}^C_{\rm relax}+\frac{\partial\hat{u}_c}{\partial\boldsymbol{c}^R}:\dot{\boldsymbol{c}}^R_{\rm relax}+\frac{\partial\hat{u}_c}{\partial\lambda}\dot{\lambda}_{\rm relax}\right).
\end{split}
\label{eq:internal_energy_bath}
\end{equation}
The second line is the work done against the entropic part of the stress, which heats the fluid immediately, and the third line is the release to the heat bath of the energy stored in the microstructure as it relaxes. For a Hookean dumbbell, for which the free energy is purely entropic, the flow term reduces to $\frac{1}{2}G(\boldsymbol{c}-\boldsymbol{\delta}):\dot{\boldsymbol{\gamma}}$, which equals $G\dot\gamma c_{xy}\geq0$ in steady simple shear: the work done against the entropic stress heats the fluid, as expected.

\section{Proofs of the propositions of Secs.~\ref{sec:thermodynamics} and~\ref{sec:model}}\label{app:proofs}

\textit{Proof of Proposition~\ref{prop:dissipation}.} Writing $\boldsymbol{A}=\partial\hat{f}_c/\partial\boldsymbol{c}^A$, which is symmetric, Eq.~\eqref{eq:entropy_generation} gives $\dot{\sigma}_A=(\rho/T)M_A\,\tr(\boldsymbol{A}\cdot\boldsymbol{c}^A\cdot\boldsymbol{A})$. Since $\boldsymbol{c}^A$ is positive definite, $\boldsymbol{A}\cdot\boldsymbol{c}^A\cdot\boldsymbol{A}=\boldsymbol{A}^\dagger\cdot\boldsymbol{c}^A\cdot\boldsymbol{A}$ is positive semidefinite, and its trace is nonnegative. For the scalar, $\dot{\sigma}_\lambda=(\rho/T)M_\lambda(\partial\hat{f}_c/\partial\lambda)^2\geq0$. \qed

\textit{Proof of Proposition~\ref{prop:stress}.} Since $\hat{f}_c=(G_R/2\rho)F$, the elastic part of \eqref{eq:stress_general} is $-G_R[\boldsymbol{c}^C\cdot\partial F/\partial\boldsymbol{c}^C+\boldsymbol{c}^R\cdot\partial F/\partial\boldsymbol{c}^R+(\partial F/\partial\lambda)\boldsymbol{g}]$. The derivatives \eqref{eq:F_derivatives} follow from $\partial D_R/\partial\boldsymbol{c}^R=\boldsymbol{\delta}-(\boldsymbol{c}^R)^{-1}$, which gives $\partial\zeta/\partial\boldsymbol{c}^R=\zeta[\boldsymbol{\delta}-(\boldsymbol{c}^R)^{-1}]$, and from $\partial\zeta/\partial\lambda=\zeta/\lambda$. By \eqref{eq:F_derivatives}, the cellular term gives $-G_R\beta(\boldsymbol{c}^C-\boldsymbol{\delta})=-G_C(\boldsymbol{c}^C-\boldsymbol{\delta})$. With \eqref{eq:g_rouge}, $(\partial F/\partial\lambda)\boldsymbol{g}=[\lambda^m-\varphi'(\zeta)\zeta](\boldsymbol{c}^R-\boldsymbol{\delta})$, and adding $\boldsymbol{c}^R\cdot\partial F/\partial\boldsymbol{c}^R$ cancels the term in $\varphi'$, leaving $-G_R\lambda^m(\boldsymbol{c}^R-\boldsymbol{\delta})$. \qed

\textit{Entropy generation of ROUGE.} Substituting \eqref{eq:F_derivatives}, \eqref{eq:mobilities}, and \eqref{eq:relaxations_rouge} into \eqref{eq:entropy_generation} gives
\begin{align}
\label{eq:entropy_generation_rouge}
\nonumber
&\dot{\sigma}_C=\frac{G_C}{2T\tau_C}\left[\tr\boldsymbol{c}^C+\tr(\boldsymbol{c}^C)^{-1}-6\right], \\
\qquad
&\dot{\sigma}_R=\frac{G_R}{2T\tau_R}\,\varphi'(\zeta)^2\left[\tr\boldsymbol{c}^R+\tr(\boldsymbol{c}^R)^{-1}-6\right], \\
\nonumber
\qquad
&\dot{\sigma}_\lambda=\frac{G_R}{2T}\,k_\lambda\left(\frac{\partial F}{\partial\lambda}\right)^2.
\end{align}
Each term is nonnegative: the brackets are nonnegative because $c_i+1/c_i\geq2$ for every positive eigenvalue, and the remaining factors are squares or nonnegative constants.

\section{Stability of the rest state}\label{app:admissibility}

\textit{Condition (C1).} Under (C1), $\varphi$ is nondecreasing, and since $\zeta=\lambda e^{D_R}\geq\lambda$ with equality only at $D_R=0$, the minimum of $\varphi$ over the deformations at fixed $\lambda$ is reached at $\boldsymbol{c}^R=\boldsymbol{\delta}$.

\textit{Local stability.} Because $\zeta/\lambda=e^{D_R}$ does not depend on $\lambda$, differentiating \eqref{eq:F_derivatives} gives
\begin{equation}
\frac{\partial^2F}{\partial\lambda^2}=\varphi''(\zeta)\left(\frac{\zeta}{\lambda}\right)^2-(m-1)\lambda^{m-2},
\end{equation}
which equals $\varphi''(1)-(m-1)$ at rest. The mixed derivatives $\partial^2F/\partial\lambda\,\partial\boldsymbol{c}^R$ are proportional to $\boldsymbol{\delta}-(\boldsymbol{c}^R)^{-1}$ and vanish at rest, so the Hessian of $F$ is block diagonal there. Its $\boldsymbol{c}^R$ block is positive definite by (C2), since near $\boldsymbol{c}^R=\boldsymbol{\delta}$ the free energy behaves as $\varphi'(1)D_R$ and $D_R$ is a positive definite quadratic form in $\boldsymbol{c}^R-\boldsymbol{\delta}$. Local stability therefore reduces to (C3).

\textit{Proof of Proposition~\ref{prop:global_min}.} The cellular part of $F$ is minimized at $\boldsymbol{c}^C=\boldsymbol{\delta}$. For the remaining part, (C1) gives $F(\lambda,D_R)\geq F(\lambda,0)$ for every $\lambda$, because $\varphi$ is nondecreasing and $\zeta\geq\lambda$. By (C4), $\partial F/\partial\lambda=\varphi'(\lambda)-\lambda^{m-1}<0$ along $D_R=0$ for $\lambda\in(0,1)$, so $F(\lambda,0)>F(1,0)$ for every $\lambda<1$. Hence $F(\lambda,D_R)\geq F(1,0)$, with equality only at the rest state. \qed

\textit{Langmuir coupling.} For the coupling \eqref{eq:langmuir_phi}, $\varphi''(\zeta)=(\zeta_c+1)\zeta_c/(\zeta_c+\zeta)^2$, so condition (C3) reads $\zeta_c/(\zeta_c+1)>m-1$, which can be satisfied only for $m<2$ and is then equivalent to Eq.~\eqref{eq:langmuir_condition}. To check (C4), we write it as $H(\lambda)<1$ with
\begin{equation}
H(\lambda)=\frac{\varphi'(\lambda)}{\lambda^{m-1}}=\frac{(\zeta_c+1)\,\lambda^{2-m}}{\zeta_c+\lambda}.
\end{equation}
Then $H(1)=1$ and $d\log H/d\lambda=(2-m)/\lambda-1/(\zeta_c+\lambda)$, which is positive for all $\lambda\leq1$ under \eqref{eq:langmuir_condition}. Hence $H$ increases on $(0,1]$ and stays below one for $\lambda<1$, so (C4) holds whenever (C3) does.

\section{Small-deformation analysis of steady shear}\label{app:steady_shear}

This appendix derives the relation~\eqref{eq:tau_aggr} between the activated mobility~\eqref{eq:k_lambda} and the shear-activation term of the t-ETV. In steady simple shear at rate $\dot\gamma$, with the rouleaux relaxation rate $\nu_R=r(\zeta)/\tau_R$, Eqs.~\eqref{eq:shear_c} give
\begin{align}
\nonumber
&c^R_{yy}=c^R_{zz}=1,
\qquad
c^R_{xy}=\frac{w}{\tau_R}\,\tau_R=w,
\qquad
c^R_{xx}=1+2w^2, \\
\qquad
&w=\frac{\mathrm{Wi}}{r(\zeta)},
\qquad
\mathrm{Wi}=\tau_R\dot\gamma.
\end{align}
Hence $\tr\boldsymbol{c}^R=3+2w^2$ and $|\boldsymbol{c}^R|=1+w^2$, so that
\begin{equation}
D_R=2w^2-\log\left(1+w^2\right)=w^2+\frac{w^4}{2}+O(w^6).
\label{eq:app_DR}
\end{equation}
This steady state is self-consistent, since $r$ depends on $\boldsymbol{c}^R$ through $\zeta=\lambda e^{D_R}$. At small $\mathrm{Wi}$, however, $D_R=O(\mathrm{Wi}^2)$, so $\zeta=\lambda+O(\mathrm{Wi}^2)$ and $r(\zeta)=r(\lambda)+O(\mathrm{Wi}^2)$, where $\lambda$ is the steady structure at the given rate. Equation~\eqref{eq:app_DR} then gives
\begin{equation}
D_R=\frac{\mathrm{Wi}^2}{r(\lambda)^2}+O(\mathrm{Wi}^4).
\end{equation}
The raw invariant $D_R$ therefore depends on the structure through $r(\lambda)$, a dependence with no counterpart in the activation term of the t-ETV. The rescaled invariant used in Eq.~\eqref{eq:k_lambda} removes it,
\begin{equation}
\tilde{D}_R=r(\zeta)^2D_R=\mathrm{Wi}^2+O(\mathrm{Wi}^4),
\end{equation}
for every $\lambda$ and every $\zeta_c$. The activation factor then becomes
\begin{equation}
\left(b\tilde{D}_R\right)^{d/2}=\left|\sqrt{b}\,\tau_R\dot\gamma\right|^d\left[1+O(\mathrm{Wi}^2)\right],
\end{equation}
which coincides at leading order with the term $|\tau_{\rm aggr}\dot\gamma|^d$ of Eq.~\eqref{eq:tetv_kinetics} if $\tau_{\rm aggr}=\sqrt{b}\,\tau_R$, which is Eq.~\eqref{eq:tau_aggr}. In the low-coverage limit, $r\equiv1$ and Eq.~\eqref{eq:app_DR} holds exactly with $w=\mathrm{Wi}$. Since the rouleaux relaxation time of the fitted model is $\tau_R=5.1\,\mathrm{ms}$, $\mathrm{Wi}\lesssim0.11$ in all the experiments considered here, and the leading-order expressions are accurate to within about $1\%$.

\section{Numerical and statistical details}\label{app:numerics}

\textit{Equations in simple shear.} For $\boldsymbol{v}=(\dot\gamma(t)\,y,0,0)$, with the convention $(\nabla\boldsymbol{v})_{ij}=\partial_iv_j$, the only nonzero component of the velocity gradient is $(\nabla\boldsymbol{v})_{yx}=\dot\gamma$. The conformation tensors keep the form $c^A_{xz}=c^A_{yz}=0$, and Eqs.~\eqref{eq:evolution_c} and~\eqref{eq:relaxations_rouge} reduce, for $A\in\{C,R\}$, to
\begin{equation}
\begin{split}
\frac{dc^A_{xx}}{dt}&=2\dot\gamma\,c^A_{xy}+\nu_A\left(1-c^A_{xx}\right),\qquad
\frac{dc^A_{xy}}{dt}=\dot\gamma\,c^A_{yy}-\nu_A\,c^A_{xy},\\
\frac{dc^A_{yy}}{dt}&=\nu_A\left(1-c^A_{yy}\right),\qquad
\frac{dc^A_{zz}}{dt}=\nu_A\left(1-c^A_{zz}\right),
\end{split}
\label{eq:shear_c}
\end{equation}
with relaxation rates $\nu_C=1/\tau_C$ and $\nu_R=r(\zeta)/\tau_R$. Since $\boldsymbol{g}:\dot{\boldsymbol{\gamma}}=-2\lambda\dot\gamma\,c^R_{xy}$, Eq.~\eqref{eq:lambda_rouge} becomes
\begin{equation}
\frac{d\lambda}{dt}=-2\lambda\dot\gamma\,c^R_{xy}-k_\lambda\frac{\partial F}{\partial\lambda},
\label{eq:shear_lambda}
\end{equation}
where $D_R=c^R_{xx}+c^R_{yy}+c^R_{zz}-\log[(c^R_{xx}c^R_{yy}-(c^R_{xy})^2)c^R_{zz}]-3$ enters through $\zeta$, $r(\zeta)$, and $k_\lambda$. Equations~\eqref{eq:shear_c} and~\eqref{eq:shear_lambda} form a system of nine ordinary differential equations. Starting from rest, the $yy$ and $zz$ components remain equal to one throughout, which leads to the reduced form of Table~\ref{tab:equations}; we integrate them nonetheless as a check of the implementation.

\textit{Integration.} We integrate the system with the LSODA method~\cite{petzold_automatic_1983} as implemented in SciPy~\cite{virtanen_scipy_2020}, supplying the Jacobian of the right-hand side, obtained symbolically with SymPy~\cite{meurer_sympy_2017}. During the parameter estimation we use relative and absolute tolerances of $10^{-6}$ and $10^{-9}$; the model curves shown in Figs.~\ref{fig:stress}--\ref{fig:reversal}, computed with the fitted parameters, use tighter tolerances of $10^{-7}$ and $10^{-10}$. In the variants with shear activation, the entries of the analytic Jacobian that contain $\tilde{D}_R^{\,d/2-1}$ diverge at $\tilde{D}_R=0$, that is, at rest; these entries are set to zero, which affects only the Newton iterations of the implicit steps and not the solution. Without activation the analytic Jacobian is finite everywhere.

\textit{Fitting variables.} The fitting variables and their bounds are listed in Table~\ref{tab:fit_variables}. All of them are fitted in logarithmic form, which keeps them positive and makes their scales comparable. The coupling constant is written as $\zeta_c=(m-1)/(2-m)+e^{\theta_\zeta}$, so that condition~\eqref{eq:langmuir_condition}, and with it (C1)--(C4), holds at every iterate. The activation constant is computed from $b=(\tau_{\rm aggr}/\tau_R)^2$ [Eq.~\eqref{eq:tau_aggr}]. The low-coverage limit is implemented exactly through Eq.~\eqref{eq:henry_F}, rather than with a large but finite $\zeta_c$. The minimization uses the trust-region reflective algorithm of SciPy~\cite{virtanen_scipy_2020}, with the Jacobian of the residuals computed by finite differences with a relative step of $10^{-3}$ and tolerances of $10^{-8}$ on the cost, the variables, and the gradient.

\begin{table}
\caption{Fitting variables and their bounds. All variables are fitted as logarithms; $\zeta_c$ is fitted through its excess over the admissibility bound \eqref{eq:langmuir_condition}. Depending on the model, either the moduli or the viscosities of the cellular and rouleaux viscoelastic modes, $\eta_p^C=G_C\tau_C$ and $\eta_R=G_R\tau_R$, are used as fitting variables (Sec.~\ref{sec:results}).}
\label{tab:fit_variables}
\begin{ruledtabular}
\begin{tabular}{lccc}
Variable & Units & Lower bound & Upper bound\\
\hline
$G_C$, $G_R$ & Pa & $10^{-2}$ & $10$\\
$\eta_p^C$, $\eta_R$ & Pa\,s & $10^{-5}$ & $10^{-1}$\\
$\eta_C$ & Pa\,s & $10^{-5}$ & $10^{-1}$\\
$\tau_C$, $\tau_R$ & s & $10^{-4}$ & $1$\\
$\tau_\lambda$ & s & $10^{-2}$ & $10^{2}$\\
$\tau_{\rm aggr}$ & s & $10^{-4}$\footnotemark[1] & $1$\\
$\zeta_c-(m-1)/(2-m)$ & -- & $10^{-3}$ & $10^{3}$\footnotemark[1]\\
$d$ & -- & $10^{-1}$ & $2$\\
\end{tabular}
\end{ruledtabular}
\footnotetext[1]{In the first, unconstrained fit, the lower bound of $\tau_{\rm aggr}$ was $10^{-3}\,\mathrm{s}$ and the upper bound of $\zeta_c-(m-1)/(2-m)$ was $10^{2}$.}
\end{table}

\textit{Covariance and model selection.} The covariance matrix of the fitting variables, linearized about the optimum, is
\begin{equation}
\mathrm{Cov}(\boldsymbol{\theta})\approx s^2\left(\mathbf{J}^\top\mathbf{J}\right)^{-1},
\qquad
s^2=\frac{2\Phi}{N-k},
\label{eq:covariance}
\end{equation}
where $\mathbf{J}$ is the Jacobian of the weighted residuals with respect to $\boldsymbol{\theta}$ and $k$ is the number of fitted variables. The Akaike information criterion is
\begin{equation}
\mathrm{AIC}=N\ln\left(\frac{2\Phi}{N}\right)+2k.
\label{eq:aic}
\end{equation}


\section*{Supplementary Material}
\noindent See the supplementary material for the parameters of all the fits of Table~\ref{tab:fits}, the cost of each experimental curve, the correlation matrices of the fitting variables, the partition of the entropy generation among cellular, rouleaux, and structural relaxation, the normalized residuals of the selected model, and a comparison of the initial parameters, fit~1, and the selected model with each experiment.

\section*{Conflicts of interest}

\noindent There are no conflicts to declare.

\section*{Data Availability Statement}
\noindent The experimental data analyzed in this study are openly available in Mendeley Data at \url{https://doi.org/10.17632/s672yykhf2.1}. The code that reproduces the fits, tables, and figures of this article will be made available at \url{https://github.com/th-martinod/rouge-blood-rheology} upon publication.




\bibliography{references}

\end{document}